\documentclass[12pt,epsf,fleqn]{article}
\usepackage{epsfig}
\usepackage[dvips]{color}
\begin{document}
\title{Higgs bosons of the NMSSM with explicit CP violation at the ILC}
\author{S.W. Ham$^{(1)}$, S.H. Kim$^{(2)}$, S.K. Oh$^{(1,2)}$, and D. Son$^{(1)}$
\\
\\
{\it $^{(1)}$ Center for High Energy Physics, Kyungpook National University}\\
{\it Daegu 702-701, Korea} \\
{\it $^{(2)}$ Department of Physics, Konkuk University, Seoul 143-701, Korea}
\\
\\
}
\date{}
\maketitle
\begin{abstract}
We study the Higgs sector of the next-to-minimal supersymmetric standard model (NMSSM)
with explicit CP violation at the one-loop level, where the radiative corrections
due to the quarks and squarks of the third generation are taken into account.
We expect that, within a reasonable region of the parameter space of the present model,
at least one of five neutral Higgs bosons may be produced at the future $e^+ e^-$
International Linear Collider (ILC) with $\sqrt{s} = 500$ GeV,
with cross section larger than 12 fb, 15 fb, and 1.5 fb, respectively, via the Higgs-strahlung process,
the $WW$ fusion process, and the $ZZ$ fusion process.
We find that the effect of the CP phase in the present model yields significant influences
upon the production cross sections of the five neutral Higgs bosons.
We also study the decay modes of the five neutral Higgs bosons
to find that their decay widths are similarly affected by the CP phase.
Some of the decay modes in the present model behave differently from those of the Standard Model.
\end{abstract}
\vfil
\eject

\section{INTRODUCTION}

These days, supersymmetry (SUSY) is one of the most important topics
for both theoretical and experimental high energy physics.
The phenomenology of SUSY has been investigated quite exhaustively recent years
with respect to the future high energy experiments at the Large Hadron Collider (LHC)
or at the International Linear Collider (ILC).
From a phenomenological point of view, SUSY must be broken in order to be realized in nature [1].
The breaking of SUSY may be accomplished by several methods,
one of which is by introducing soft SUSY breaking terms.

The exact form of the soft SUSY breaking terms are not precisely determined.
They may differ from one model to another.
In particular, the coefficients of these SUSY breaking terms
and their associated terms may either be real or complex.
The complex phases that may arise from these soft SUSY breaking terms
in a supersymmetric extension of the standard model (SM) can be the source of CP violation,
through the mixing between the CP even and the CP odd states,
since there are at least two Higgs doublets in the model
in order to give masses to the up-quark sector and the down-quark sector separately [2].

It is already known that, at the tree level, neither explicit nor spontaneous CP violation is possible
in the Higgs sector of the minimal supersymmetric standard model (MSSM), which has just two Higgs doublets,
because any complex phases in its Higgs sector can always be eliminated by rotating the Higgs fields.
Even at the one-loop level, spontaneous CP violation is disfavored
because it requires a very light neutral Higgs boson, which has already been ruled out by experiments.
On the other hand, explicit CP violation is viable in the MSSM at the one-loop level
since the radiative corrections due to the loops of relevant particles,
such as quarks and squarks, yield the mixing between the CP even and the CP odd neutral Higgs bosons [3-10].

The next-to-minimal supersymmetric standard model (NMSSM), with an additional Higgs singlet
besides two Higgs doublets in its Higgs sector, may not accommodate spontaneous CP violation
at the tree level because of vacuum stability [11].
However, the scenario of explicit CP violation is possible for the NMSSM at the tree level.
By assuming the degeneracy of the stop quark masses in the Higgs sector of the NMSSM,
large explicit CP violation may be realized as the vacuum expectation value (VEV) of the neutral Higgs singlet
approaches to the electroweak scale [12].
The tree-level Higgs potential of the NMSSM may have only one physical CP violating phase,
either by redefining the phases of the three Higgs fields or
by applying two CP-odd tadpole minimum conditions in the explicit CP violation scenario.

In previous articles, we have analyzed the phenomenology of the explicit CP violation
in the Higgs sector of the NMSSM at the 1-loop level,
where the radiative corrections due to quarks and squarks of the third generation are taken into account [13].
There can be several complex phases which allow the explicit CP violation
in the Higgs sector of the NMSSM at the one-loop level:
Two complex phases may arise from the stop and sbottom quark masses
that are associated with quadratic coupling $\lambda$ and the trilinear SUSY breaking parameters,
$A_t$ and $A_b$.
Additional complex phases from the chargino and neutralino sectors may also contribute
to the CP mixing among the scalar and pseudoscalar Higgs bosons.

The presence of explicit CP violation may affect the Higgs phenomenology such as the masses of
the neutral Higgs bosons, their productions and decays.
In high-energy $e^+e^-$ collisions, the dominant production channels for neutral Higgs bosons
are the Higgs-strahlung process, the $WW$ fusion process, and the $ZZ$ fusion process.
Within the context of explicit CP violation, the prospects of discovering the neutral Higgs bosons
in high-energy $e^+e^-$ collisions have been studied in a general two-Higgs doublet model [14] and
in the MSSM [15,16].
In particular, in the MSSM with explicit CP violation, where the CP mixing among the CP even and CP odd states
is maximized,
the production cross sections for the neutral Higgs bosons in $e^+ e^-$ collisions
at $\sqrt{s}$ = 500 and 800 GeV are calculated [16].
Meanwhile, the decays of the Higgs bosons in the MSSM with explicit CP violation have
also been studied in the literature [5,10].

In this article, we study the Higgs sector of the NMSSM at the one-loop level
in the explicit CP violation scenario,
where  the quarks and squarks of the third generation are taken into account.
The two CP-odd tadpole minimum conditions are obtained at the one-loop level.
There appear three physical phases as the possible sources of explicit CP violation in the NMSSM.
We establish and explore a reasonable region in the parameter space of the present model,
including the CP phases, in order to examine how the upper bound on the masses of the neutral Higgs bosons
are dependent on the parameters.

Further, the production cross sections, and decay rates as well, of the neutral Higgs bosons are calculated
for the future $e^+e^-$ ILC with $\sqrt{s} = 500$ GeV, within the established parameter region,
via the Higgs-strahlung process, the $WW$ fusion process, and the $ZZ$ fusion process.
We study the effects of the CP phases upon the production cross sections of these neutral Higgs bosons
and their decay rates.
We also obtain the lower bound on the production cross sections of these neutral Higgs bosons
within the established parameter region, which allow us to expect that the neutral sector of the NMSSM
with explicit CP violation may be tested at the future ILC with $\sqrt{s} = 500$ GeV.

\section{THE HIGGS SECTOR}

The Higgs sector of the NMSSM consists of two Higgs doublet fields $H_1 = (H_1^0, H^-)$,
$H_2 = (H^+, H_2^0)$, and a neutral Higgs singlet field $N$ [17,18].
Their hypercharges are -1/2, 1/2, and 0, respectively.
For the fermion matter fields in this model, only the quarks of the third generation are taken into account.
Then, the superpotential of the model may be expressed as
\begin{equation}
W = h_t Q H_2 t_R^c + h_b Q H_1 b_R^c + \lambda N H_1^T \epsilon H_2 - {k \over 3} N^3 \ ,
\end{equation}
where $h_q (q=t,b)$ are the Yukawa coupling coefficients,
$Q$ is the chiral superfield containing the left-handed $t,b$ quarks,
$q_R^c$ ($q=t,b$) are the charge conjugates of the right-handed $t,b$ quarks,
and $\epsilon$ is an antisymmetric $2 \times 2$ matrix with $\epsilon_{12}=1$.
Note that all coupling coefficients are dimensionless.

As is well known, the parameter $\lambda$ is introduced into the NMSSM
to avoid the so-called $\mu$-problem in the MSSM, since $\lambda x$ corresponds
to the $\mu$ parameter in the superpotetial of the MSSM, where $x$ is the VEV of $N$.
The global U(1) Peccei-Quinn symmetry is explicitly broken
by the presence of the cubic term proportional to $k$ in the superpotential.
Without the $k$-term, a massless pseudoscalar Higgs boson would appear
beacuse the Peccei-Quinn symmetry is recovered at the tree level.
Thus, the two dimensionless parameters $\lambda$ and $k$ are essential in the NMSSM.

The Higgs potential at the tree level may decompose into three parts as [18]
\[
    V_0 = V_D + V_F + V_{\rm S} \ ,
\]
where
\begin{eqnarray}
V_D & = & {g_2^2 \over 8} (H_1^{\dag} \vec\sigma H_1 + H_2^{\dag} \vec\sigma H_2)^2
+ {g_1^2\over 8}(|H_2|^2 - |H_1|^2)^2  \ , \cr
V_F & = & |\lambda|^2[(|H_1|^2+|H_2|^2)|N|^2+|H_1^T \epsilon H_2|^2] + |k|^2|N|^4
-(\lambda k^*H_1^T \epsilon H_2 N^{*2}+ {\rm H.c.}) \ , \cr
V_{\rm S} & = & m_{H_1}^2|H_1|^2 + m_{H_2}^2|H_2|^2 + m_N^2|N|^2
- (\lambda A_{\lambda} H_1^T \epsilon H_2 N + {1\over 3} k A_k N^3 + {\rm H.c.})\ ,
\end{eqnarray}
with $g_1$ and $g_2$ being the U(1) and SU(2) gauge coupling constants, respectively,
and $\vec \sigma = (\sigma^1, \sigma^2, \sigma^3)$ are the Pauli matrices.
The soft SUSY breaking $V_{\rm S}$ has three soft masses $m_{H_1}$, $m_{H_2}$, and $m_N$,
and two additional parameters $A_{\lambda}$ and $A_k$, both of mass dimension.

In general, $\lambda$, $k$, $A_{\lambda}$, and $A_k$ in $V_0$ can be complex.
Among them, $\lambda A_{\lambda}$ and $k A_k$ can be adjusted to be real and positive
by redefining the phases of $H_1^T \epsilon H_2$ and $N$.
Thus, at the tree level, $k e^{i \phi_k}$ and $\lambda  e^{i \phi_\lambda}$ can
in general be complex.
Consequently, the tree-level Higgs potential may have at most only one physical phase,
which may be chosen to be $\phi = \phi_\lambda -\phi_k$ in $\lambda k^*$.

The radiative corrections at the one-loop level to the above tree-level Higgs potential,
obtained by employing the effective potential method, is given as [19]
\begin{equation}
V_1  = \sum_{l} {n_l {\cal M}_l^4 \over 64 \pi^2}
\left [
\log {{\cal M}_l^2 \over \Lambda^2} - {3 \over 2}
\right ]  \ ,
\end{equation}
where the subscript $l$ stands for the quarks and squarks of the third generation,
$n_q = -12$ and $n_{\tilde q} = 6$ are respectively the degrees of freedom for quarks and squarks,
and $\Lambda$ is the renormalization scale in the modified minimal subtraction scheme.
In the above expression, note that quarks enter with a negative sign
while squarks enter with a positive sign, and ${\cal M}_l^2$ depends
on the neutral Higgs fields, where the tree-level masses of the relevant quark
and squarks should be used as input.
The radiatively corrected Higgs potential at the one-loop level is given by $V = V_0 + V_1$.

After the electroweak symmetry breaking, neutral Higgs fields $H_1^0$, $H_2^0$, and $N$
may acquire complex VEVs $v_1$, $v_2 e^{i \theta}$, and $x e^{i \delta}$,
respectively, where $v_1$ is adjusted to be real.
The tree-level masses of the quarks of the third generation are given as $m_t = h_t v \sin \beta$
and $m_b = h_b v \cos \beta$, where $\tan\beta = v_2/v_1$ and $v = \sqrt{v_1^2 + v_2^2} = 175$ GeV.

The masses of the squarks of the third generation are
\begin{eqnarray}
m_{{\tilde t}_1, {\tilde t}_2}^2 & = & m_Q^2 + m_t^2 \mp m_t
\sqrt{A_t^2 + \lambda^2 x^2 \cot^2 \beta + 2 \lambda A_t x \cot \beta \cos \phi_t} \ , \cr
m_{{\tilde b}_1, {\tilde b}_2}^2 & = & m_Q^2 + m_b^2 \mp m_b
\sqrt{A_b^2 + \lambda^2 x^2 \tan^2 \beta + 2 \lambda A_b x \tan \beta \cos \phi_b} \ ,
\end{eqnarray}
where $m_Q$ is the soft SUSY breaking mass, and $A_t$ and $A_b$ are
the trilinear SUSY breaking parameters of the mass dimension.
Two CP phases $\phi_t$ and $\phi_b$ are given by $\phi_t = \theta + \delta + \phi_{\lambda} + \phi_{A_t}$
and $\phi_b = \theta + \delta + \phi_{\lambda} + \phi_{A_b}$, where $\phi_{\lambda}$
and $\phi_{A_q} (q=t,b)$ come from the complex phases of $\lambda$ and $A_q (q=t,b)$,
respectively.
Note that the masses of squarks of the third generation are different from the masses
of the quarks of the third generation, due to the SUSY breaking terms, which account
in general for the mass splitting between an ordinary particle and its superpartner
depending on the SUSY breaking scale (1-2 TeV).

For simplicity, we reexpress the Higgs multiplets via a unitary transformation as
\begin{eqnarray}
\begin{array}{lll}
        H_1 & = & \left ( \begin{array}{c}
          v_1 + S_1 + i \sin \beta A   \cr
          \sin \beta C^{+ *}
  \end{array} \right )  \ ,  \cr
        H_2 & = & \left ( \begin{array}{c}
          \cos \beta C^+           \cr
          (v_2 + S_2 + i \cos \beta A) e^{i \theta}
  \end{array} \right )   \ ,  \cr
        N & = & \left ( \begin{array}{c}
          (x + X + i Y) e^{i \delta}
  \end{array} \right )   \ ,
\end{array}
\end{eqnarray}
where $S_1$, $S_2$ and $X$ are the scalar Higgs fields, $A$ and $Y$ are the pseudoscalar Higgs fields,
and $C^+$ is the charged Higgs field.
Note that the phases are explicitly shown in the expression.

As we allow CP violation in the Higgs sector of the NMSSM at the one-loop level,
there are several complex phases: $\phi_k$, $\phi_{\lambda}$, $\phi_{A_{\lambda}}$,
$\phi_{A_k}$, $\phi_{A_t}$, and $\phi_{A_b}$.
These phases arise respectively from $k$, $\lambda$, $A_{\lambda}$, $A_k$,
$A_t$, and $A_b$.
In addition to these phases, we have two more complex phases, namely, $\theta$ and $\delta$.
Thus, we have eight complex phases in the radiatively corrected Higgs potential.
The two CP-odd tadpole minimum conditions, with respect to $A$ and $Y$, are obtained as
\begin{eqnarray}
0 & = & k x \sin \phi + A_{\lambda} \sin \phi_1
+ {3 h_t^2 \over 16 \pi^2} A_t \sin \phi_t f(m_{{\tilde t}_1}^2, m_{{\tilde t}_2}^2) \cr
& &\mbox{}+ {3 h_b^2 \over 16 \pi^2} A_b \sin \phi_b f(m_{{\tilde b}_1}^2, m_{{\tilde b}_2}^2) \ , \cr
0 & = &\mbox{} - 2 k v^2 \lambda x \sin 2 \beta \sin \phi + A_{\lambda} \lambda v^2 \sin 2 \beta \sin \phi_1
- 2 k A_k x^2 \sin \phi_2 \cr
& &\mbox{} + {3 h_t^2 \over 16 \pi^2} A_t \lambda v^2 \sin 2 \beta \sin \phi_t
f(m_{{\tilde t}_1}^2, m_{{\tilde t}_2}^2) \cr
& &\mbox{}+ {3 h_b^2 \over 16 \pi^2} A_b \lambda v^2 \sin 2 \beta \sin \phi_b
f(m_{{\tilde b}_1}^2, m_{{\tilde b}_2}^2) \ ,
\end{eqnarray}
where the dimensionless function $f(m_x^2, m_y^2)$ is defined by
\begin{equation}
 f(m_x^2, m_y^2) = {1 \over (m_y^2 - m_x^2)} \left[  m_x^2 \log {m_x^2 \over \Lambda^2} - m_y^2
\log {m_y^2 \over \Lambda^2} \right] + 1 \ ,
\end{equation}
and $\phi = \phi_{\lambda} - \phi_k - 2 \delta + \theta$,
$\phi_1 = \phi_{\lambda} + \phi_{A_{\lambda}} + \theta + \delta$,
and $\phi_2 = \phi_k + \phi_{A_k} + 3 \delta$,
with $\phi_t = \theta + \delta + \phi_{\lambda} + \phi_{A_t}$
and $\phi_b = \theta + \delta + \phi_{\lambda} + \phi_{A_b}$.
Note that $\phi$ here at the one-loop level is different from the phase $\phi$
previously defined at the tree level.
After imposing these two CP-odd tadpole minimum conditions,
we are left with only three phases $\phi$ and $\phi_t$ and $\phi_b$
in the radiatively corrected Higgs potential $V$ at the one-loop level.
These three phases are the physically free parameters.

For the masses of the five neutral Higgs fields, we differentiate $V$
with respect to the neutral Higgs fields,
in order to obtain the symmetric $5\times 5$ mass matrix $M$ for the neutral Higgs bosons.
The matrix elements are calculated in the ($S_1, S_2, A, X, Y$)-basis as
\begin{eqnarray}
M_{11} & = & M_{11}^q + (m_Z \cos \beta)^2 + m_A^2 \sin^2 \beta , \cr
M_{22} & = & M_{22}^q + (m_Z \sin \beta)^2 + m_A^2 \cos^2 \beta , \cr
M_{33} & = & M_{33}^q + m_A^2  \ , \cr
M_{44} & = & M_{44}^q + {v^2 \over 4 x^2} m_A^2 \sin^2 2 \beta - {v^2 \over 2} \lambda k \sin 2 \beta \cos \phi + (2 k x)^2
- k x A_k \cos \phi_2 \ , \cr
M_{55} & = & M_{55}^q + {v^2 \over 4 x^2} m_A^2 \sin^2 2 \beta + {3 v^2 \over 2} \lambda k \sin 2 \beta \cos \phi
+ 3 k x A_k \cos \phi_2 \ , \cr
M_{12} & = & M_{12}^q + (2 \lambda^2 v^2 - m_Z^2 - m_A^2) \sin \beta \cos \beta \ , \cr
M_{13} & = & M_{13}^q  \ , \cr
M_{14} & = & M_{14}^q - {v \over x} m_A^2 \sin^2 \beta \cos \beta - v \lambda k x \sin \beta \cos \phi
+ 2 v \lambda^2 x \cos \beta \ , \cr
M_{15} & = & M_{15}^q + - 3 \lambda k v x \sin \beta \sin \phi \ , \cr
M_{23} & = & M_{23}^q  \ , \cr
M_{24} & = & M_{24}^q - {v \over x} m_A^2 \sin \beta \cos^2 \beta - v \lambda k x \cos \beta \cos \phi
+ 2 v \lambda^2 x \sin \beta \ , \cr
M_{25} & = & M_{25}^q - 3 \lambda k v x \cos \beta \sin \phi \ , \cr
M_{34} & = & M_{34}^q + \lambda k v x \sin \phi \ , \cr
M_{35} & = & M_{35}^q + {v \over x} m_A^2 \sin \beta \cos \beta - 3 v \lambda k x \cos \phi \ , \cr
M_{45} & = & M_{45}^q + 2 \lambda k v^2 \sin 2 \beta \sin \phi \ ,
\end{eqnarray}
where $m_A^2$ is introduced for convenience as
\begin{eqnarray}
m_A^2 & = & {\lambda x (A_{\lambda} \cos \phi_1 + k x \cos \phi) \over \sin \beta \cos \beta}
+ {3 m_t^2 A_t \lambda x \cos \phi_t \over 16 \pi^2 v^2 \sin^3 \beta \cos \beta}
f (m_{{\tilde t}_1}^2, m_{{\tilde t}_2}^2) \cr
& &\mbox{} + {3 m_b^2 A_b \lambda x \cos \phi_b \over 16 \pi^2 v^2 \cos^3 \beta \sin \beta}
f (m_{{\tilde b}_1}^2, m_{{\tilde b}_2}^2) \ ,
\end{eqnarray}
and $M_{ij}^q = M_{ij}^t + M_{ij}^b$ are the matrix elements due to the contributions of the third generations.
Explicitly, the contributions due to top and stop quarks are given as
\begin{eqnarray}
M_{11}^t & = & {3 m_t^4 \lambda^2 x^2 \Delta_{{\tilde t}_1}^2 \over 8 \pi^2  v^2 \sin^2 \beta}
{g(m_{\tilde{t}_1}^2, \ m_{\tilde{t}_2}^2) \over (m_{\tilde{t}_2}^2 - m_{\tilde{t}_1}^2)^2}   \ , \cr
M_{22}^t & = & {3 m_t^4 A_t^2 \Delta_{{\tilde t}_2}^2 \over 8 \pi^2  v^2 \sin^2 \beta}
{g(m_{\tilde{t}_1}^2, \ m_{\tilde{t}_2}^2) \over (m_{\tilde{t}_2}^2 - m_{\tilde{t}_1}^2)^2}
+ {3 m_t^4 A_t \Delta_{{\tilde t}_2} \over 4 \pi^2 v^2 \sin^2 \beta}
{\log (m_{\tilde{t}_2}^2 / m_{\tilde{t}_1}^2)  \over (m_{\tilde{t}_2}^2 - m_{\tilde{t}_1}^2)} \cr
& &\mbox{} + {3 m_t^4 \over 8 \pi^2 v^2 \sin^2 \beta}
\log \left ( {m_{\tilde{t}_1}^2  m_{\tilde{t}_2}^2 \over m_t^4} \right ) \ , \cr
M_{33}^t & = & {3 m_t^4 \lambda^2 x^2 A_t^2 \sin^2 \phi_t \over 8 \pi^2 v^2 \sin^4 \beta}
{g(m_{\tilde{t}_1}^2, \ m_{\tilde{t}_2}^2) \over (m_{\tilde{t}_2}^2 - m_{\tilde{t}_1}^2 )^2}  \ , \cr
M_{44}^t & = & {3 m_t^4 \lambda^2 \Delta_{{\tilde t}_1}^2 \over 8 \pi^2 \tan^2 \beta}
{g(m_{\tilde{t}_1}^2, \ m_{\tilde{t}_2}^2) \over (m_{\tilde{t}_2}^2 - m_{\tilde{t}_1}^2 )^2}  \ , \cr
M_{55}^t & = & {3 m_t^4 \lambda^2 A_t^2 \sin^2 \phi_t \over 8 \pi^2 \tan^2 \beta}
{g(m_{\tilde{t}_1}^2, \ m_{\tilde{t}_2}^2) \over (m_{\tilde{t}_2}^2 - m_{\tilde{t}_1}^2 )^2}  \ ,  \cr
M_{12}^t & = & {3 m_t^4 \lambda x A_t \Delta_{{\tilde t}_1} \Delta_{{\tilde t}_2} \over 8 \pi^2 v^2 \sin^2 \beta}
{g(m_{\tilde{t}_1}^2, \ m_{\tilde{t}_2}^2) \over (m_{\tilde{t}_2}^2 - m_{\tilde{t}_1}^2)^2}
+ {3 m_t^4 \lambda x \Delta_{{\tilde t}_1} \over 8 \pi^2 v^2 \sin^2 \beta}
{\log (m_{\tilde{t}_2}^2 / m_{\tilde{t}_1}^2) \over (m_{\tilde{t}_2}^2 - m_{\tilde{t}_1}^2)}  \ , \cr
M_{13}^t & = & \mbox{} - {3 m_t^4 \lambda^2 x^2 A_t \Delta_{{\tilde t}_1} \sin \phi_t \over 8 \pi^2 v^2 \sin^3 \beta}
{g(m_{\tilde{t}_1}^2, \ m_{\tilde{t}_2}^2) \over (m_{\tilde{t}_2}^2 - m_{\tilde{t}_1}^2)^2 } \ , \cr
M_{14}^t & = & {3 m_t^4 \lambda^2 x \Delta_{{\tilde t}_1}^2 \over 8 \pi^2 v \sin \beta \tan \beta}
{g(m_{\tilde{t}_1}^2, \ m_{\tilde{t}_2}^2) \over (m_{\tilde{t}_2}^2 - m_{\tilde{t}_1}^2)^2 }
- {3 m_t^2 \lambda^2 x \cot \beta \over 8 \pi^2 v \sin \beta} f(m_{\tilde{t}_1}^2, \ m_{\tilde{t}_2}^2)  , \cr
M_{15}^t & = &\mbox{} - {3 m_t^4 \lambda^2 x A_t \Delta_{{\tilde t}_1} \sin \phi_t \over 8 \pi^2 v \sin \beta \tan \beta}
{g(m_{\tilde{t}_1}^2, \ m_{\tilde{t}_2}^2) \over (m_{\tilde{t}_2}^2 - m_{\tilde{t}_1}^2)^2 } \ , \cr
M_{23}^t & = & \mbox{} - {3 m_t^4 \lambda x A_t^2 \Delta_{{\tilde t}_2} \sin \phi_t \over 8 \pi^2 v^2 \sin^3 \beta}
{g(m_{\tilde{t}_1}^2, \ m_{\tilde{t}_2}^2) \over (m_{\tilde{t}_2}^2 - m_{\tilde{t}_1}^2)^2 }
- {3 m_t^4 \lambda x A_t \sin \phi_t \over 8 \pi^2 v^2 \sin^3 \beta}
{\log (m_{\tilde{t}_2}^2 / m_{\tilde{t}_1}^2) \over (m_{\tilde{t}_2}^2 - m_{\tilde{t}_1}^2)}  \ , \cr
M_{24}^t & = & {3 m_t^4 \lambda A_t \Delta_{{\tilde t}_1} \Delta_{{\tilde t}_2} \over 8 \pi^2 v \sin \beta \tan \beta}
{g(m_{\tilde{t}_1}^2, \ m_{\tilde{t}_2}^2) \over (m_{\tilde{t}_2}^2 - m_{\tilde{t}_1}^2)^2}
 + {3 m_t^4 \lambda \Delta_{{\tilde t}_1} \over 8 \pi^2 v \sin \beta \tan \beta}
{\log (m_{\tilde{t}_2}^2 / m_{\tilde{t}_1}^2) \over (m_{\tilde{t}_2}^2 - m_{\tilde{t}_1}^2) }  , \cr
M_{25}^t & = & \mbox{} - {3 m_t^4 \lambda A_t^2 \Delta_{{\tilde t}_2} \sin \phi_t \over 8 \pi^2 v \sin \beta \tan \beta}
{g(m_{\tilde{t}_1}^2, \ m_{\tilde{t}_2}^2) \over (m_{\tilde{t}_2}^2 - m_{\tilde{t}_1}^2)^2 }
- {3 m_t^4 \lambda A_t \sin \phi_t \over 8 \pi^2 v \sin \beta \tan \beta}
{ \log (m_{\tilde{t}_2}^2 / m_{\tilde{t}_1}^2) \over (m_{\tilde{t}_2}^2 - m_{\tilde{t}_1}^2)}  \ , \cr
M_{34}^t & = & \mbox{} - {3 m_t^4 \lambda^2 x A_t \Delta_{{\tilde t}_1} \sin \phi_t \over 8 \pi^2 v \sin^2 \beta \tan \beta}
{g(m_{\tilde{t}_1}^2, \ m_{\tilde{t}_2}^2) \over (m_{\tilde{t}_2}^2 - m_{\tilde{t}_1}^2)^2 } \ , \cr
M_{35}^t & = & \mbox{} {3 m_t^4 \lambda^2 x A_t^2 \sin^2 \phi_t \over 8 \pi^2 v \sin^2 \beta \tan \beta}
{g(m_{\tilde{t}_1}^2, \ m_{\tilde{t}_2}^2) \over (m_{\tilde{t}_2}^2 - m_{\tilde{t}_1}^2)^2 }  \ , \cr
M_{45}^t & = & \mbox{} - {3 m_t^4 \lambda^2 A_t \Delta_{{\tilde t}_1} \sin \phi_t \over 8 \pi^2 \tan^2 \beta}
{ g(m_{\tilde{t}_1}^2, \ m_{\tilde{t}_2}^2) \over (m_{\tilde{t}_2}^2 - m_{\tilde{t}_1}^2)^2 } \ ,
\end{eqnarray}
and the contributions due to bottom and sbottom quarks are given as
\begin{eqnarray}
M_{11}^b & = & {3 m_b^4 A_b^2 \Delta_{{\tilde b}_1}^2 \over 8 \pi^2  v^2 \cos^2 \beta}
{g(m_{\tilde{b}_1}^2, \ m_{\tilde{b}_2}^2) \over (m_{\tilde{b}_2}^2 - m_{\tilde{b}_1}^2)^2}
+ {3 m_b^4 A_b \Delta_{{\tilde b}_1} \over 4 \pi^2 v^2 \cos^2 \beta}
{\log (m_{\tilde{b}_2}^2 / m_{\tilde{b}_1}^2)  \over (m_{\tilde{b}_2}^2 - m_{\tilde{b}_1}^2)} \cr
& &\mbox{} + {3 m_b^4 \over 8 \pi^2 v^2 \cos^2 \beta}
\log \left ( {m_{\tilde{b}_1}^2  m_{\tilde{b}_2}^2 \over m_b^4} \right ) \ , \cr
M_{22}^b & = & {3 m_b^4 \lambda^2 x^2 \Delta_{{\tilde b}_2}^2 \over 8 \pi^2  v^2 \cos^2 \beta}
{g(m_{\tilde{b}_1}^2, \ m_{\tilde{b}_2}^2) \over (m_{\tilde{b}_2}^2 - m_{\tilde{b}_1}^2)^2}   \ , \cr
M_{33}^b & = & {3 m_b^4 \lambda^2 x^2 A_b^2 \sin^2 \phi_b \over 8 \pi^2 v^2 \cos^4 \beta}
{g(m_{\tilde{b}_1}^2, \ m_{\tilde{b}_2}^2) \over (m_{\tilde{b}_2}^2 - m_{\tilde{b}_1}^2 )^2} \ , \cr
M_{44}^b & = & {3 m_b^4 \lambda^2 \Delta_{{\tilde b}_2}^2 \over 8 \pi^2 \cot^2 \beta}
{g(m_{\tilde{b}_1}^2, \ m_{\tilde{b}_2}^2) \over (m_{\tilde{b}_2}^2 - m_{\tilde{b}_1}^2 )^2}  \ , \cr
M_{55}^b & = & {3 m_b^4 \lambda^2 A_b^2 \sin^2 \phi_b \over 8 \pi^2 \cot^2 \beta}
{g(m_{\tilde{b}_1}^2, \ m_{\tilde{b}_2}^2) \over (m_{\tilde{b}_2}^2 - m_{\tilde{b}_1}^2 )^2} \ ,  \cr
M_{12}^b & = & {3 m_b^4 \lambda x A_b \Delta_{{\tilde b}_1} \Delta_{{\tilde b}_2} \over 8 \pi^2 v^2 \cos^2 \beta}
{g(m_{{\tilde b}_1}^2, \ m_{{\tilde b}_2}^2) \over (m_{{\tilde b}_2}^2 - m_{{\tilde b}_1}^2)^2}
+ {3 m_b^4 \lambda x \Delta_{{\tilde b}_2} \over 8 \pi^2 v^2 \cos^2 \beta}
{\log (m_{{\tilde b}_2}^2 / m_{{\tilde b}_1}^2)  \over (m_{{\tilde b}_2}^2 - m_{{\tilde b}_1}^2)} \ , \cr
M_{13}^b & = & \mbox{} - {3 m_b^4 \lambda x A_b^2 \Delta_{{\tilde b}_1} \sin \phi_b \over 8 \pi^2 v^2 \cos^3 \beta}
{g(m_{\tilde{b}_1}^2, \ m_{\tilde{b}_2}^2) \over (m_{\tilde{b}_2}^2 - m_{\tilde{b}_1}^2)^2 }
- {3 m_b^4 \lambda x A_b \sin \phi_b \over 8 \pi^2 v^2 \cos^3 \beta}
{\log (m_{\tilde{b}_2}^2 / m_{\tilde{b}_1}^2) \over (m_{\tilde{b}_2}^2 - m_{\tilde{b}_1}^2)}  \ , \cr
M_{14}^b & = & {3 m_b^4 \lambda A_b \Delta_{{\tilde b}_1} \Delta_{{\tilde b}_2} \over 8 \pi^2 v \cos \beta \cot \beta}
{g(m_{\tilde{b}_1}^2, \ m_{\tilde{b}_2}^2) \over (m_{\tilde{b}_2}^2 - m_{\tilde{b}_1}^2)^2}
 + {3 m_b^4 \lambda \Delta_{{\tilde b}_2} \over 8 \pi^2 v \cos \beta \cot \beta}
{\log (m_{\tilde{b}_2}^2 / m_{\tilde{b}_1}^2) \over (m_{\tilde{b}_2}^2 - m_{\tilde{b}_1}^2) }  , \cr
M_{15}^b & = & \mbox{} - {3 m_b^4 \lambda A_b^2 \Delta_{{\tilde b}_1} \sin \phi_b \over 8 \pi^2 v \cos \beta \cot \beta}
{g(m_{\tilde{b}_1}^2, \ m_{\tilde{b}_2}^2) \over (m_{\tilde{b}_2}^2 - m_{\tilde{b}_1}^2)^2 }
- {3 m_b^4 \lambda A_b \sin \phi_b \over 8 \pi^2 v \cos \beta \cot \beta}
{ \log (m_{\tilde{b}_2}^2 / m_{\tilde{b}_1}^2) \over (m_{\tilde{b}_2}^2 - m_{\tilde{b}_1}^2)}  \ , \cr
M_{23}^b & = & \mbox{} - {3 m_b^4 \lambda^2 x^2 A_b \Delta_{{\tilde b}_2} \sin \phi_b \over 8 \pi^2 v^2 \cos^3 \beta}
{g(m_{\tilde{b}_1}^2, \ m_{\tilde{b}_2}^2) \over (m_{\tilde{b}_2}^2 - m_{\tilde{b}_1}^2)^2 } \ , \cr
M_{24}^b & = & {3 m_b^4 \lambda^2 x \Delta_{{\tilde b}_2}^2 \over 8 \pi^2 v \cos \beta \cot \beta}
{g(m_{\tilde{b}_1}^2, \ m_{\tilde{b}_2}^2) \over (m_{\tilde{b}_2}^2 - m_{\tilde{b}_1}^2)^2 }
- {3 m_b^2 \lambda^2 x \tan \beta \over 8 \pi^2 v \cos \beta} f(m_{\tilde{b}_1}^2, \ m_{\tilde{b}_2}^2)  , \cr
M_{25}^b & = &\mbox{} - {3 m_b^4 \lambda^2 x A_b \Delta_{{\tilde b}_2} \sin \phi_b \over 8 \pi^2 v \cos \beta \cot \beta}
{g(m_{\tilde{b}_1}^2, \ m_{\tilde{b}_2}^2) \over (m_{\tilde{b}_2}^2 - m_{\tilde{b}_1}^2)^2 } \ , \cr
M_{34}^b & = & \mbox{} - {3 m_b^4 \lambda^2 x A_b \Delta_{{\tilde b}_2} \sin \phi_b \over 8 \pi^2 v \cos^2 \beta \cot \beta}
{g(m_{\tilde{b}_1}^2, \ m_{\tilde{b}_2}^2) \over (m_{\tilde{b}_2}^2 - m_{\tilde{b}_1}^2)^2 }   \ , \cr
M_{35}^b & = & \mbox{} {3 m_b^4 \lambda^2 x A_b^2 \sin^2 \phi_b \over 8 \pi^2 v \cos^2 \beta \cot \beta}
{g(m_{\tilde{b}_1}^2, \ m_{\tilde{b}_2}^2) \over (m_{\tilde{b}_2}^2 - m_{\tilde{b}_1}^2)^2 }  \ , \cr
M_{45}^b & = & \mbox{} - {3 m_b^4 \lambda^2 A_b \Delta_{{\tilde b}_2} \sin \phi_b \over 8 \pi^2 \cot^2 \beta}
{ g(m_{\tilde{b}_1}^2, \ m_{\tilde{b}_2}^2) \over (m_{\tilde{b}_2}^2 - m_{\tilde{b}_1}^2)^2 } \ ,
\end{eqnarray}
where
\begin{eqnarray}
 \Delta_{{\tilde t}_1} &=& A_t \cos \phi_t + \lambda x \cot \beta  \  , \cr
 \Delta_{{\tilde t}_2} & = & A_t + \lambda x \cot \beta \cos \phi_t \ , \cr
 \Delta_{{\tilde b}_1} & = & A_t + \lambda x \tan \beta \cos \phi_b \ , \cr
 \Delta_{{\tilde b}_2} &=& A_b \cos \phi_t + \lambda x \tan \beta  \  ,
\end{eqnarray}
and the dimensionless function $g(m_x^2,m_y^2)$ is defined by
\begin{equation}
 g(m_x^2,m_y^2) = {m_y^2 + m_x^2 \over m_x^2 - m_y^2} \log {m_y^2 \over m_x^2} + 2 \ .
\end{equation}
Note that since these matrix elements are calculated at the one-loop level,
they naturally contain $f(m_{{\tilde t}_1}^2, m_{{\tilde t}_2}^2)$ or $f(m_{{\tilde b}_1}^2, m_{{\tilde b}_2}^2)$.
In the above expressions, we use the two CP-odd tadpole minimum conditions to eliminate most of them.

The matrix elements that are responsible for the scalar-pseudoscalar mixing
are $M_{13}$, $M_{15}$, $M_{23}$, $M_{25}$, $M_{34}$, and $M_{45}$.
The physical neutral Higgs bosons and their squared masses are given respectively
by the eigenvectors and the eigenvalues, denoted as $m^2_{h_i}$ ($i$ = 1-5), of the mass matrix.
We sort these five neutral Higgs bosons in the increasing order of their masses such that,
for example,  $m^2_{h_1}$ is the smallest eigenvalue and $h_1$ is the lightest neutral Higgs boson.

Finally, the charged Higgs boson mass at the tree level is obtained as
\begin{equation}
m_C^2 = m_W^2 - \lambda^2 v^2 + {2 \lambda x \over \sin 2 \beta} (A_{\lambda} \cos \phi_1 + k x \cos \phi) \ .
\end{equation}
where $m_W$ is the $W$ boson mass.
Note that, unlike the case of the MSSM, the tree-level mass of the charged Higgs boson
may be either heavier or lighter than $W$ boson, according to whether the second term is
smaller than the third term.
Also note that the charged Higgs boson mass contains $\cos \phi$,
even though there might be no CP mixing in the charged Higgs sector.

\section{NUMERICAL ANALYSES}

For numerical analyses, we take $\sin^2 \theta_W = 0.23$ for weak-mixing angle,
$G_F = 1.166 \times 10^{- 5}$ for the Fermi coupling constant,
$\Gamma_Z = 2.48$ GeV for the total decay width of $Z$ boson,
$m_Z = 91.187$ GeV for the mass of $Z$ boson, $m_W = 80.423$ GeV for the mass of $W$ boson,
$m_t = 175$ GeV for the mass of top quark, and $m_b = 4$ GeV for the mass of bottom quark.
In the NMSSM with explicit CP violation, we have as many as thirteen free parameters:
$\Lambda$, $\phi$, $\phi_t$, $\phi_b$, $m_Q$, $\tan \beta$, $\lambda$, $k$,
$A_k$, $x$, $A_t$, $A_b$, and $A_{\lambda}$.
Among them, we set $\Lambda = 300$ GeV, $\phi_t = \phi_b$, and $A_t = A_b$ for simplicity.
Then, we are left with ten free parameters $\phi$, $\phi_t$, $m_Q$, $\tan \beta$, $\lambda$,
$k$, $A_k$, $x$, $A_t$, and $A_{\lambda}$.
However, we will use $m_A$ in Eq. 9 instead of $A_{\lambda}$,
since it is the lighter pseudoscalar Higgs boson mass in CP-conserving limit.

We explore a reasonable region of the parameter space of the model, whose boundaries
are set as $0 < \phi, \phi_t < \pi$, $2 <\tan \beta < 30$, $0 < \lambda, k < 0.8$,
$0 < m_Q, A_k, x, m_A < 1000$ GeV, and $0 < A_t < 2000$ GeV, using a random-number generating function.
The number of points we explore in this region of the parameter space is $2 \times 10^5$.
Among those $2 \times 10^5$ points randomly distributed in the parameter space,
only 14571 points are physically allowed.
At these 14571 points of the parameter space, we consider the neutral Higgs productions
at the ILC, the future high-energy $e^+e^-$ colliding machine, with the proposed center of mass energy of 500 GeV.

\subsection{HIGGS MASS}

The mass of the lightest neutral Higgs boson, $m_{h_1}$, is calculated
at each of the physically allowed points we explore in the prescribed region,
and the masses of heavier neutral Higgs bosons as well.
The result is shown in Fig. 1, where $m_{h_1}$ is plotted against $\tan \beta$.
One can find that the upper bound on the lightest neutral Higgs boson mass increases
as $\tan \beta$ increases, and it is nearly saturated at about 150 GeV as $\tan \beta$ approaches to 30.
The behavior of the upper bound on the  lightest neutral Higgs boson mass in the NMSSM
with explicit CP violation is somewhat similar to that of the MSSM with no CP violation.
It may be contrasted against the case of no CP violation in the Higgs sector of the NMSSM,
where the upper bound on the lightest neutral Higgs boson mass becomes maximal
at around $\tan \beta \sim 2.5$ and decreases slightly as $\tan\beta$ further increases,
for a wide region in the parameter space, mainly because of the Higgs singlet effects.

\subsection{HIGGS PRODUCTION}

The most important channels for the production of neutral Higgs bosons in high energy $e^+e^-$ collisions are:
the Higgs-strahlung process $e^+e^- \rightarrow Z h_i$,
the $WW$ fusion process $e^+e^- \rightarrow {\bar \nu}_e \nu_e h_i $, and
the $ZZ$ fusion process $e^+e^- \rightarrow e^+ e^- h_i$ ($i$ = 1-5).
We denote the production cross sections for each of these processes
as $\sigma_i^H (m_{h_i})$, $\sigma_i^W (m_{h_i})$, and $\sigma_i^Z (m_{h_i})$, respectively.
These production cross sections in each channel are related to the production cross section
for the SM Higgs boson in the corresponding channel as [20]
\[
    \sigma_i^H (m_{h_i}) = R_i^2 \sigma_{\rm SM}^H (m_{h_i}) \ ,  \
    \sigma_i^W (m_{h_i}) = R_i^2 \sigma_{\rm SM}^W (m_{h_i}) \ , \
    \sigma_i^Z (m_{h_i}) = R_i^2 \sigma_{\rm SM}^Z (m_{h_i})  \ ,
\]
where $R_i$ ($i$ = 1-5) are defined as
\begin{equation}
R_i = \cos \beta O_{1i} + \sin \beta O_{2i}  \ ,
\end{equation}
with $O_{ij}$ being the elements of the orthogonal matrix that diagonalizes
the $5 \times 5$ mass matrix of the neutral Higgs bosons.
They satisfy $0 \le R^2_i \le 1$ ($i$ = 1-5) and the orthogonal condition of $\sum_{i = 1}^5 R_i^2 = 1$.
Note that they represent collectively all relevant parameters of the NMSSM
at the one-loop level, including the CP violating phases.

Let us first consider the Higgs-strahlung process for the neutral Higgs productions.
At each physically allowed point, we calculate the mass of the participating neutral Higgs boson,
and its production cross section via the Higgs-strahlung process.
We repeat this job for all five neutral Higgs bosons, to obtain $\sigma_i^H (m_{h_i})$ ($i$ = 1-5)
at each of 14571 points.
Now, we denote by $\sigma_0 (hZZ)$ the largest production cross section among them, namely,
\begin{equation}
\sigma_0 (hZZ) = {\rm max} [\sigma_1^H, \sigma_2^H, \sigma_3^H, \sigma_4^H, \sigma_5^H] \ .
\end{equation}
Thus, at every physically allowed point of the parameter space of the NMSSM, $\sigma_0 (hZZ)$ is obtained.
The meaning of $\sigma_0 (hZZ)$ is that, at a given point of the parameter space,
at least one of the five neutral Higgs bosons in the NMSSM may be produced
via the Higgs-strahlung process with that cross section.

In Fig 2a, we plot $\sigma_0 (hZZ)$ against $m_{h_1}$ for the future high energy $e^+e^-$ collider
with $\sqrt{s} =500$ GeV.
The dashed curve in the figure represents the production cross section of the SM Higgs boson
via the Higgs-strahlung process [20,21].
Among $\sigma_0 (hZZ)$, we may select the smallest one and denote it as $\sigma^H_0$.
The significance of $\sigma^H_0$ is quite clear; at least one of the five neutral Higgs bosons
in the NMSSM may be produced via the Higgs-strahlung process
with a cross section larger than $\sigma^H_0$, regardless of the parameter values.
In this sense, it may be regarded as the absolute minimum of the production cross section
of a neutral Higgs boson in the NMSSM via the Higgs-strahlung process, with explicit CP violation
at the one-loop level.
From Fig. 2a, one can see that $\sigma^H_0 \sim 12$ fb.

Next, we consider the neutral Higgs productions via the $WW$ fusion process and the $ZZ$ fusion process.
For the $WW$ fusion process, we determine $\sigma_0 (h\nu{\bar \nu})$ and $\sigma^W_0$,
which are respectively analogous to $\sigma_0 (hZZ)$ and $\sigma^H_0$ for the Higgs-strahlung process.
We show the result for $\sigma_0 (h\nu{\bar \nu})$ in Fig. 2b, plotted against  $m_{h_1}$,
and we obtain $\sigma^W_0 \sim 15$ fb.
Also, for the $ZZ$ fusion process, we determine $\sigma_0 (he^+e^-)$ and $\sigma^Z_0$,
which are respectively analogous to $\sigma_0 (hZZ)$ and $\sigma^H_0$ for the Higgs-strahlung process.
We show the result for $\sigma_0 (he^+e^-)$  in Fig. 2c, plotted against  $m_{h_1}$,
and we obtain $\sigma^Z_0 \sim 1.5$ fb.
The dashed curves in Figs. 2b and 2c represent respectively the production cross sections
of the SM Higgs boson via the $WW$ fusion and the $ZZ$ fusion processes.

We find that, interestingly, all of $\sigma^H_0 \sim 12$ fb, $\sigma^W_0 \sim 15$ fb,
and $\sigma^Z_0 \sim 1.5$ fb occur at an identical point of the parameter spce,
where the values of the parameters are $\phi \approx \phi_t \approx \pi/6$, $\tan \beta = 20$,
$m_Q =766$ GeV, $\lambda = 0.088$, $k = 0.119$, $A_k = 86$ GeV, $x = 699$ GeV, $A_t = 1350$ GeV,
and $m_A = 139$ GeV.
We may regard this point of the parameter space with these values as the most difficult point
for the neutral Higgs productions in $e^+ e^-$ collisions with $\sqrt{s} = 500$ GeV.
We would like to investigate the region around this point in more detail.

Let us study a narrow region defined as $0 \le \phi_t =\phi \le \pi$, $0 \le \tan\beta \le 20$,
$m_Q = 766$ GeV, $\lambda = 0.088$, $k = 0.12$, $A_k = 86$ GeV, $x = 699$ GeV, $A_t = 1350$ GeV,
and $m_A = 139$ GeV.
It is in fact a two-dimensional subspace in the ($\phi, \tan \beta$)-plane.
We calculate $\sigma_0 (hZZ)$ in this restricted region.
The result is shown in Fig. 3a.
Likewise, the result for $\sigma_0 (h \nu {\bar \nu})$ and $\sigma_0 (h e^+e^-)$
are respectively shown in Figs. 3b and 3c.
One can see that indeed $\sigma^H_0 \sim 12$ fb, $\sigma^W_0 \sim 15$ fb,
and $\sigma^Z_0 \sim 1.5$ fb at the point $(\phi, \tan \beta) = (\pi/6, 20)$
in the ($\phi, \tan \beta$)-plane, indicated by the arrow, in Figs. 3a, 3b, and 3c,
respectively, for the neutral Higgs productions in $e^+ e^-$ collisions with $\sqrt{s} = 500$ GeV.

Now, we study the effect of the CP phases on the neutral Higgs productions in $e^+ e^-$ collisions
with $\sqrt{s} = 500$ GeV.
The parameters are set as $0 \le \phi_t =\phi \le \pi$, $\tan\beta = 20$, $m_Q = 766$ GeV,
$\lambda = 0.088$, $k = 0.12$, $A_k = 86$ GeV, $x = 699$ GeV, $A_t = 1350$ GeV, and $m_A = 139$ GeV.
Thus, we are investigating a one-dimensional subspace of the line in $\phi$ direction.

In this one-dimensional subspace of the parameter space, we calculate the masses
of the five neutral Higgs bosons in the present model, in order to see the dependencies
of the neutral Higgs masses upon the CP phase.
The results are shown in Fig. 4a, where the five neutral Higgs boson masses are plotted
as functions of $\phi$.
One can notice in Fig. 4a that the five neutral Higgs boson masses may vary up to about 5 GeV,
as the CP phase changes within the given range.

The coefficients $R_i$ ($i$ = 1-5), which relate the neutral Higgs production cross sections
in the present model with the production cross sections for the SM Higgs boson,
also depend on the parameters of the present model.
We plot in Fig. 4b $R_i$ as functions of $\phi$.
All $R_i$ fluctuate quite significantly as $\phi$ changes within the given range.
These wild behaviors of $R_i$ indicate that the production cross sections of the neutral Higgs bosons
would also fluctuate largely against the CP phase.
The production cross sections of the five neutral Higgs bosons are displayed as functions of $\phi$,
via the Higgs-strahlung, $WW$ fusion, and $ZZ$ fusion processes, respectively, in Figs. 4c, 4d, and 4e.
As might be expected, all of the neutral Higgs production cross sections fluctuate largely against $\phi$.
In fact, the patterns of fluctuation in the neutral Higgs productions is very similar to the fluctuation patterns
in $R_i$.
This is mainly because the dependencies of the masses of the neutral Higgs bosons upon the CP phase
are relatively negligible as compared to the dependencies of $R_i$ upon the CP phase.

The three figures, Figs. 4c, 4d, and 4e, are useful to determine $\sigma_0 (hZZ)$,
$\sigma_0 (h\nu{\bar \nu})$, and $\sigma_0 (he^+e^-)$, as well as $\sigma^H_0$,
$\sigma^W_0$, and $\sigma^Z_0$, respectively.
Consider for example the neutral Higgs productions via the Higgs-strahlung process
in Fig. 4c, where all of $\sigma_1^H$, $\sigma_2^H$, $\sigma_3^H$, $\sigma_4^H$,
and $\sigma_5^H$ are shown.
Since $\sigma_0 (hZZ)$ is defined
as $\sigma_0 (hZZ) = {\rm max} [\sigma_1^H, \sigma_2^H, \sigma_3^H, \sigma_4^H, \sigma_5^H]$,
we may identify the envelope of the five curves in Fig. 4c as $\sigma_0 (hZZ)$.
Thus, in Fig. 4c, the curve for $\sigma_0 (hZZ)$ is somewhat W-shaped.
One may notice that the minimum of $\sigma_0 (hZZ)$, which is defined as $\sigma^H_0$,
occurs at $\phi \sim 6/\pi$ with $\sigma_0^H \sim 12$ fb.
Likewise, one may also notice in Figs. 4d and 4e respectively that $\sigma_0^W \sim 15$ fb
for the neutral Higgs productions via the $WW$ fusion process and $\sigma_0^Z \sim 1.5$ fb
for the neutral Higgs productions via the $ZZ$ fusion process.
The results obtained from Figs. 4c, 4d, and 4e are consistent with the results
from Figs. 2a, 2b, and 2c, respectively.

\subsection{HIGGS DECAY}

Now, let us study the decay modes of the five neutral Higgs bosons $h_j$ ($j$ = 1-5)
in the NMSSM with explicit CP violation.
The partial decay width of a neutral Higgs boson $h_j$ into a pair of down or up quarks is given as
\begin{eqnarray}
\Gamma (h_j \to d {\bar d}) & = & {C_f g_2^2 m_d^2 m_{h_j} \over 32 \pi m_W^2}
\sqrt{1- \tau_d}
\left[{O_{j1}^2 \over \cos^2 \beta} (1 - \tau_d)
+ \tan^2 \beta O_{j3}^3 \right ] \ , \cr
\Gamma (h_j\to u {\bar u}) & = & {C_f g_2^2 m_u^2 m_{h_j} \over 32 \pi m_W^2}
\sqrt{1-\tau_u }
\left[{O_{j2}^2 \over \sin^2 \beta} (1 - \tau_u)
+ \cot^2 \beta O_{j3}^3 \right ] \ ,
\end{eqnarray}
where $C_f = 3$ is the color factor of quarks, and $\tau_q = 4 m_q^2 /m_{h_j}^2$ ($q = d,u$).
The color factor of leptons is 1.
The partial decay width of a neutral Higgs boson $h_j$ into a pair of charged leptons
is the same as $\Gamma (h_j \to d {\bar d})$ with $C_f = 1$ for the color factor of leptons.

The partial decay width of $h_j$ into a pair of gauge bosons may be obtained from that
of the SM Higgs boson, through a relation given by
\begin{equation}
\Gamma (h_j \to VV) = R_j^2 \ \Gamma_{\rm SM} (h_j \to VV) \ ,
\end{equation}
where $\Gamma_{\rm SM} (h_j \to VV)$ is the decay width of the SM Higgs boson into a pair of gauge bosons.
The produced gauge bosons may be either real or virtual.

The partial decay width of $h_j$ into a pair of photons is given as
\begin{eqnarray}
\Gamma (h_j \to \gamma \gamma) & = & {\alpha^2 m_{h_j}^3 \over 576 \pi^3}
\left|C_f e_b^2 {O_{j1} \over \cos \beta} S_b (\tau_b) + C_f e_t^2 {O_{j2} \over \sin \beta} S_t (\tau_t)
- R_i S_W (\tau_W) \right|^2 \cr
& &\mbox{} + {\alpha^2 m_{h_j}^3 \over 576 \pi^3}
\left|C_f e_b^2 \tan \beta O_{j3} P_b (\tau_b) + C_f e_t^2 \cot \beta O_{j3} P_t (\tau_t) \right|^2 \ ,
\end{eqnarray}
where $e_t$ and $e_b$ are respectively the electric charges of top and bottom quarks,
$\tau_W = 4 m_W^2 /m_{h_j}^2$, and $S_q (\tau_q)$, $P_q (\tau_q)$ ($q = t,b$),
and $S_W (\tau_W)$ are respectively the form factors of the scalar, pseudoscalar,
and gauge bosons, given as [10,22]
\begin{eqnarray}
&& S_q (\tau_q) =  {3 \over 2} \tau_q (1 + (1-\tau_q) f(\tau_q)) \ , \cr
&& P_q (\tau_q) =  {3 \over 2} \tau_q f(\tau_q) \ , \cr
&& S_W (\tau_W) = - [2 + 3 \tau_W + 3 \tau_W (2 - \tau_W) f(\tau_W) ] \ ,
\end{eqnarray}
where the function $f$ is defined as
\begin{eqnarray}
f(\tau) = \left \{
\begin{array}{cl}
{\rm arcsin}^2(1/\sqrt{\tau})    & \qquad \tau \geq 1 \ , \cr
-{1 \over 4} \left[ \log \left( {1+\sqrt{1+ \tau} \over 1 - \sqrt{1- \tau}} \right) -i \pi \right]^2
& \qquad \tau < 1 \ .
\end{array}\right.
\end{eqnarray}

Lastly, the partial decay width of $h_j$ into a pair of gluons is given as
\begin{eqnarray}
\Gamma (h_j \to gg) & = & {g_2^2 m_{h_j}^3 (\alpha_s (m_{h_j}))^2 \over 288 \pi^3 m_W^2}
\left|{O_{j1} \over \cos \beta} S_b (\tau_b) + {O_{j2} \over \sin \beta} S_t (\tau_t) \right|^2 \cr
&&\mbox{} \times \left[1+ S_{N_F} {\alpha_s (m_{h_j}) \over \pi} \right] \cr
& &\mbox{} + {g_2^2 m_{h_j}^3 (\alpha_s (m_{h_j}))^2 \over 288 \pi^3 m_W^2}
\left| \tan \beta O_{j3} P_b (\tau_b) + \cot \beta O_{j3} P_t (\tau_t) \right|^2 \cr
&&\mbox{} \times \left[1+ P_{N_F} {\alpha_s (m_{h_j}) \over \pi} \right] \ .
\end{eqnarray}
where $\alpha_s (m_{h_j})$ is the coupling coefficient of the strong interactions,
$N_F$ is the number of the quark flavors lighter than $h_j$, and
\begin{equation}
S_{N_F} = {95 \over 4} - {7 \over 6} N_F \ , \qquad  P_{N_F} = {97 \over 4} - {7 \over 6} N_F \ ,
\end{equation}
for $m^2_{h_j} \ll 4 m_q^2$.

Now, we calculate the total decay width of the neutral Higgs bosons in the NMSSM
with explicit CP violation, in the parameter range set for Fig. 4a, namely,
$0 \le \phi \le \pi$ ($\phi=\phi_t=\phi_b$), $\tan\beta = 20$, $m_Q = 766$ GeV,
$\lambda = 0.088$, $k = 0.12$, $A_k = 86$ GeV, $x = 699$ GeV, $A_t = 1350$ GeV, and $m_A = 139$ GeV.
Let us remind the reader that the neutral Higgs bosons are most difficult to be discovered
in this parameter range.
The masses of all neutral Higgs bosons lie within the range of 120 to 160 GeV, as shown in Fig. 4a.
Thus, those decay modes listed above are the dominant channels for the Higgs boson research.
We assume that the total decay widths of $h_j$ is given by
\begin{eqnarray}
\Gamma(h_j) & = & \Gamma (h_j \to b{\bar b}) + \Gamma (h_j \to \tau^+\tau^-) + \Gamma (h_j \to c {\bar c})
+ \Gamma (h_j \to s {\bar s}) \cr
& &\mbox{} + \Gamma (h_j \to W^+W^-) + \Gamma (h_j \to ZZ) + \Gamma (h_j \to \gamma \gamma) + \Gamma (h_j \to gg) \ .
\end{eqnarray}
In Fig. 5a, the total decay widths of the five neutral Higgs bosons are plotted
as functions of the phase $\phi$.
Note that the total decay widths fluctuate widely against $\phi$.
The results shown in Fig. 5a may be compared to the case of the SM,
where the total decay width of the SM Higgs boson is obtained between 0.003535 GeV
and 0.08613 GeV as the SM Higgs boson mass increases from 120 GeV to 160 GeV.
On the other hand, the total decay width of $h_1$ in the present model may be
as large as about 0.63 GeV, according to Fig. 5a.

In Figs. 5b, 5c, 5d, 5e, and 5f, the branching ratios of each of the five neutral Higgs bosons
are shown as functions of $\phi$, in the same parameter range set for Fig. 4a or Fig. 5a.
One may easily observe, in these figures, that for the whole range of $0 \le \phi \le \pi$,
$\Gamma (h_j \to b {\bar b})$ is dominant over other partial decay widths,
for all five neutral Higgs bosons, except for Fig. 5e, where the branching ratio of $h_4$
into $W^+W^-$ is larger than that of $h_4$ into $b {\bar b}$, in a narrow range near $\phi \sim 2\pi/3$.
The decay widths of the five neutral Higgs bosons into $\tau^+ \tau^-$ are also important,
except $h_4$, in the present model, at least for larger $\tan\beta$.
Note that we set $\tan \beta = 20$ in our numerical calculations.

One may also observe that for the most part of the $\phi$ range the branching ratio of $h_j$
into $s {\bar s}$ is larger than that of into $c {\bar c}$, for all five neutral Higgs bosons except $h_4$.
We have a large branching ratio of $h_j$ into $s {\bar s}$, mainly because we set $\tan\beta =20$,
since the decay width of a neutral Higgs boson into a pair of down-type quarks, as well as that
into a pair of charged leptons, increases as $\tan \beta$ increases.
This observation is interesting as compared to the case of the SM, where the branching ratio
of the SM Higgs boson into $c{\bar c}$ is larger than that into $s {\bar s}$
because the coupling strength of the SM Higgs boson with a fermion is proportional to the mass of the fermion.

\section{CONCLUSIONS}

We have studied the Higgs sector of the NMSSM with explicit CP violation,
at the one-loop level where the radiative corrections due to the quarks and squarks
of the third generation are taken into account.
The CP-odd tadpole minimum conditions at the one-loop level are analytically obtained,
and the mass matrix for the five neutral Higgs bosons are calculated.
Unlike the CP conserving case in the Higgs sector of the NMSSM,
the upper bound on the mass of the lightest neutral Higgs boson
in the present model is found to increase as $\tan \beta$ increases,
in the parameter region set as $0 < \phi, \phi_t < \pi$, $0 < \lambda, k < 0.8$,
$0 < m_Q, A_k, x, m_A < 1000$ GeV, and $0 < A_t < 2000$ GeV.

The production cross sections of the five neutral Higgs bosons,
via the Higgs-strahlung process, the $WW$ fusion process,
and the $ZZ$ fusion process, are calculated in the parameter region set
as $0 < \phi, \phi_t < \pi$, $0 < \lambda, k < 0.8$, $0 < m_Q, A_k, x, m_A < 1000$ GeV,
and $0 < A_t < 2000$ GeV.
It is found that, in this parameter range, at least one of the five neutral Higgs bosons
may be produced at the ILC with $\sqrt{s} = 500$ GeV with cross sections larger
than 12, 15, and 1.5 fb, respectively, via the Higgs-strahlung process,
the $WW$ fusion process, and the $ZZ$ fusion process.
These numbers suggest that it is possible for the ILC with $\sqrt{s} = 500$ GeV
to test the present model.

Further, the effects of the CP phase on the productions of the five neutral Higgs bosons
via the Higgs-strahlung process, the $WW$ fusion process, and the $ZZ$ fusion process,
are studied for the parameter region set as $0 \le \phi \le \pi$ ($\phi=\phi_t=\phi_b$),
$\tan\beta = 20$, $m_Q = 766$ GeV, $\lambda = 0.088$, $k = 0.12$, $A_k = 86$ GeV,
$x = 699$ GeV, $A_t = 1350$ GeV, and $m_A = 139$ GeV.
This is a one-dimensional subspace in $\phi$ direction of the parameter space
where it is found that the production cross sections
for at least one of the five neutral Higgs bosons are minimum.
In this one-dimensional parameter space of $0 \le \phi \le \pi$,
the cross sections of the five neutral Higgs bosons are found to be seriously dependent
on $\phi$, regardless of via which process they are produced.
Their large fluctuations against $\phi$ suggest that the explicit CP violation in the NMSSM
through $\phi$ is indeed very crucial for the study of the Higgs sector in the present model.

The decay widths of the five neutral Higgs bosons into various particles are also studied
in the parameter region set as $0 \le \phi \le \pi$ ($\phi=\phi_t=\phi_b$), $\tan\beta = 20$,
$m_Q = 766$ GeV, $\lambda = 0.088$, $k = 0.12$, $A_k = 86$ GeV, $x = 699$ GeV, $A_t = 1350$ GeV,
and $m_A = 139$ GeV.
It is found that the present model predicts somewhat different patterns of neutral Higgs boson decays
from those of the SM or those of the NMSSM with no CP violation.
Whereas the branching ratio of the SM Higgs boson into $c{\bar c}$ is larger than that into $s{\bar s}$,
the branching ratio of $h_j$ in the present into $s{\bar s}$ is larger than that into $c{\bar c}$,
for the most part of the $\phi$ range, for all five neutral Higgs bosons except $h_4$.

In conclusion, the scenario of explicit CP violation in the Higgs sector of the NMSSM
exhibits interesting behaviors of the present model and thus is worthwhile for further study.

\vskip 0.3 in
\noindent
{\large {\bf ACKNOWLEDGMENTS}}
\vskip 0.2 in
\noindent
This research was supported by KOSEF through CHEP.
The authors would like to acknowledge the support from KISTI
(Korea Institute of Science and Technology Information) under
"The Strategic Supercomputing Support Program" with Dr. Kihyeon Cho as the technical supporter.
The use of the computing system of the Supercomputing Center is also greatly appreciated.

\vskip 0.3 in



\vfil\eject
{\noindent\bf FIGURE CAPTIONS}

\vskip 0.25 in
\noindent
Fig. 1: The mass of the lightest neutral Higgs boson is plotted against $\tan \beta$,
for $0 < \phi, \phi_t < \pi$, $0 < \lambda, k < 0.8$, $0 < m_Q, A_k, x, m_A < 1000$ GeV,
and $0 < A_t < 2000$ GeV.

\vskip 0.25 in
\noindent
Fig. 2(a): The plot of $\sigma_0 (hZZ)$ against $m_{h_1}$ in $e^+e^-$ collisions
with $\sqrt{s} =500$ GeV, for $0 < \phi, \phi_t < \pi$, $2 <\tan \beta < 30$,
$0 < \lambda, k < 0.8$, $0 < m_Q, A_k, x, m_A < 1000$ GeV, and $0 < A_t < 2000$ GeV.
The dashed curve represents the production cross section of the SM Higgs boson
via the Higgs-strahlung process.
The minimum value of $\sigma_0 (hZZ)$ is about 12 fb, represented by a point
in the lower right corner at about $m_{h_1} \sim 127$ GeV.

\vskip 0.25 in
\noindent
Fig. 2(b): The plot of $\sigma_0 (h \nu {\bar \nu})$ against $m_{h_1}$ in $e^+e^-$ collisions
with $\sqrt{s} =500$ GeV, for the same parameter ranges as Fig. 2a.
The dashed curve represents the production cross section of the SM Higgs boson
via the $WW$ fusion process.
The minimum value of $\sigma_0 (h \nu {\bar \nu})$ is about 15 fb, represented
by a point in the lower right corner at about $m_{h_1} \sim 127$ GeV.

\vskip 0.25 in
\noindent
Fig. 2(c): The plot of $\sigma_0 (h e^+e^-)$ against $m_{h_1}$ in $e^+e^-$ collisions
with $\sqrt{s} =500$ GeV, for the same parameter ranges as Fig. 2a.
The dashed curve represents the production cross section of the SM Higgs boson
via the $ZZ$ fusion process.
The minimum value of $\sigma_0 (h e^+e^-)$ is about 1.5 fb, represented
by a point in the lower right corner at about $m_{h_1} \sim 127$ GeV.

\vskip 0.25 in
\noindent
Fig. 3(a): The lego plot of $\sigma_0 (hZZ)$ in the ($\tan \beta, \phi$)-plane
for the Higgs production via the Higgs-strahlung process in $e^+e^-$ collisions
with $\sqrt{s} =500$ GeV, for $\phi_t =\phi$, $m_Q = 766$ GeV, $\lambda = 0.088$,
$k = 0.12$, $A_k = 86$ GeV, $x = 699$ GeV, $A_t = 1350$ GeV, and $m_A = 139$ GeV.
The minimum value of $\sigma_0 (hZZ)$ is about 12 fb, which occurs at a dip
in the forefront of the ($\tan \beta, \phi$)-plane at ($\tan \beta, \phi$) = ($20, 3/\pi$),
indicated by an arrow.

\vskip 0.25 in
\noindent
Fig. 3(b): The lego plot of $\sigma_0 (h\nu{\bar \nu})$ in the ($\tan \beta, \phi$)-plane
for the Higgs production via the $WW$ fusion process in $e^+e^-$ collisions with $\sqrt{s} =500$ GeV,
for the same parameters as Fig. 3a.
The minimum value of $\sigma_0 (h\nu{\bar \nu})$ is about 15 fb, which occurs
at the same point of the ($\tan \beta, \phi$)-plane as Fig. 3a.

\vskip 0.25 in
\noindent
Fig. 3(c): The lego plot of $\sigma_0 (h e^+e^-)$ in the ($\tan \beta, \phi$)-plane
for the Higgs production via the $ZZ$ fusion process in $e^+e^-$ collisions with $\sqrt{s} =500$ GeV,
for the same parameters as Fig. 3a.
The minimum value of $\sigma_0 (h e^+e^-)$ is about 1.5 fb, which occurs at the same point
of the ($\tan \beta, \phi$)-plane as Fig. 3a.

\vskip 0.25 in
\noindent
Fig. 4(a): The masses of the five neutral Higgs bosons are plotted as functions of $\phi$,
where the parameters are set as $\phi_t =\phi$, $\tan \beta =20$, $m_Q = 766$ GeV,
$\lambda = 0.088$, $k = 0.12$, $A_k = 86$ GeV, $x = 699$ GeV, $A_t = 1350$ GeV, and $m_A = 139$ GeV.

\vskip 0.25 in
\noindent
Fig. 4(b): $R_i$ ($i$ = 1-5) as functions of $\phi$, for the same parameters as Fig. 4a.

\vskip 0.25 in
\noindent
Fig. 4(c): The production cross sections of $h_i$ ($i$ = 1-5) via the Higgs-strahlung process
in $e^+e^-$ collisions with $\sqrt{s} =500$ GeV, as functions of $\phi$, for the same parameters as Fig. 4a.

\vskip 0.25 in
\noindent
Fig. 4(d): The production cross sections of $h_i$ ($i$ = 1-5) via the $WW$ fusion process
in $e^+e^-$ collisions with $\sqrt{s} =500$ GeV, as functions of $\phi$, for the same parameters as Fig. 4a.

\vskip 0.25 in
\noindent
Fig. 4(e): The production cross sections of $h_i$ ($i$ = 1-5) via the $ZZ$ fusion process
in $e^+e^-$ collisions with $\sqrt{s} =500$ GeV, as functions of $\phi$, for the same parameters as Fig. 4a.

\vskip 0.25 in
\noindent
Fig. 5(a): The total decay widths of the five neutral Higgs bosons as functions of $\phi$,
for the same parameters as Fig. 4a.

\vskip 0.25 in
\noindent
Fig. 5(b): The branching ratios of $h_1$ as functions of $\phi$, for the same parameters as Fig. 4a.

\vskip 0.25 in
\noindent
Fig. 5(c): The branching ratios of $h_2$ as functions of $\phi$, for the same parameters as Fig. 4a.

\vskip 0.25 in
\noindent
Fig. 5(d): The branching ratios of $h_3$ as functions of $\phi$, for the same parameters as Fig. 4a.

\vskip 0.25 in
\noindent
Fig. 5(e): The branching ratios of $h_4$ as functions of $\phi$, for the same parameters as Fig. 4a.

\vskip 0.25 in
\noindent
Fig. 5(f): The branching ratios of $h_5$ as functions of $\phi$, for the same parameters as Fig. 4a.

\vfil\eject

\renewcommand\thefigure{1}
\begin{figure}[t]
\begin{center}
\includegraphics[scale=0.6]{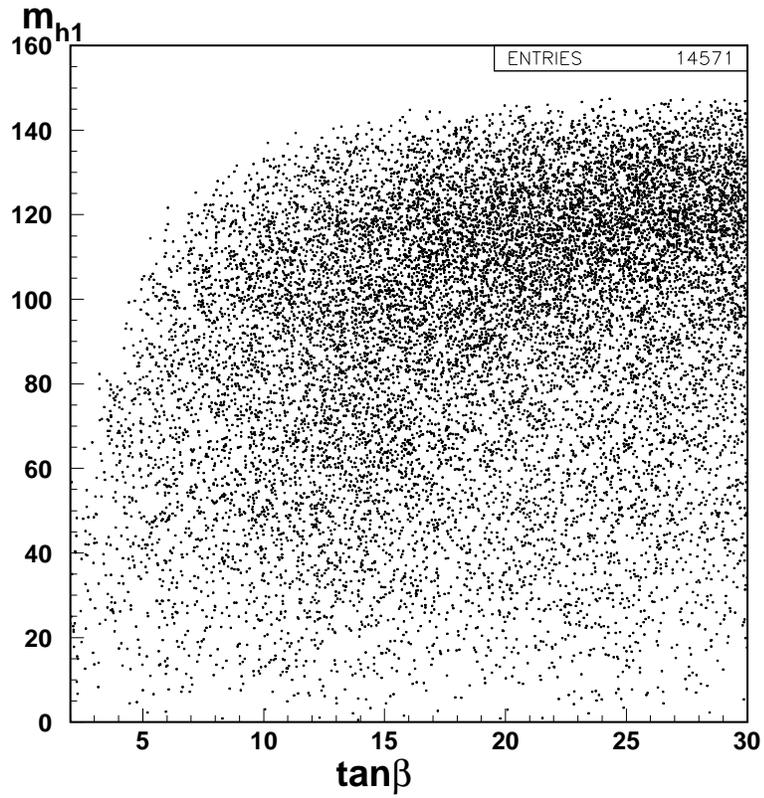}
\caption[plot]{The mass of the lightest neutral Higgs boson is plotted against $\tan \beta$,
for $0 < \phi, \phi_t < \pi$, $0 < \lambda, k < 0.8$, $0 < m_Q, A_k, x, m_A < 1000$ GeV,
and $0 < A_t < 2000$ GeV.}
\end{center}
\end{figure}

\renewcommand\thefigure{2a}
\begin{figure}[t]
\begin{center}
\includegraphics[scale=0.6]{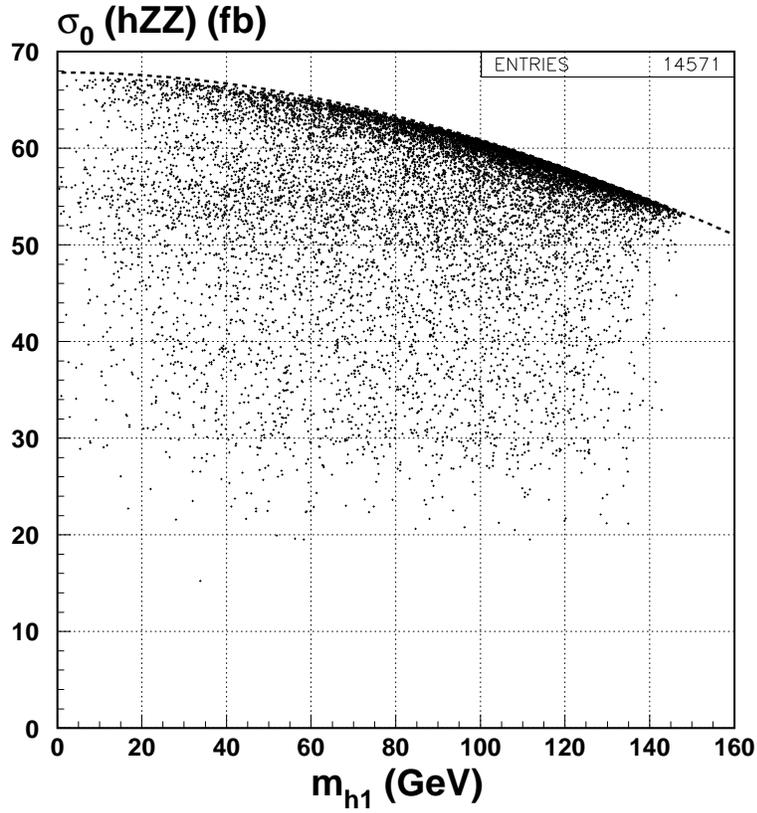}
\caption[plot]{The plot of $\sigma_0 (hZZ)$ against $m_{h_1}$ in $e^+e^-$ collisions
with $\sqrt{s} =500$ GeV, for $0 < \phi, \phi_t < \pi$, $2 <\tan \beta < 30$,
$0 < \lambda, k < 0.8$, $0 < m_Q, A_k, x, m_A < 1000$ GeV, and $0 < A_t < 2000$ GeV.
The dashed curve represents the production cross section of the SM Higgs boson
via the Higgs-strahlung process.
The minimum value of $\sigma_0 (hZZ)$ is about 12 fb, represented by a point
in the lower right corner at about $m_{h_1} \sim 127$ GeV.}
\end{center}
\end{figure}

\renewcommand\thefigure{2b}
\begin{figure}[t]
\begin{center}
\includegraphics[scale=0.6]{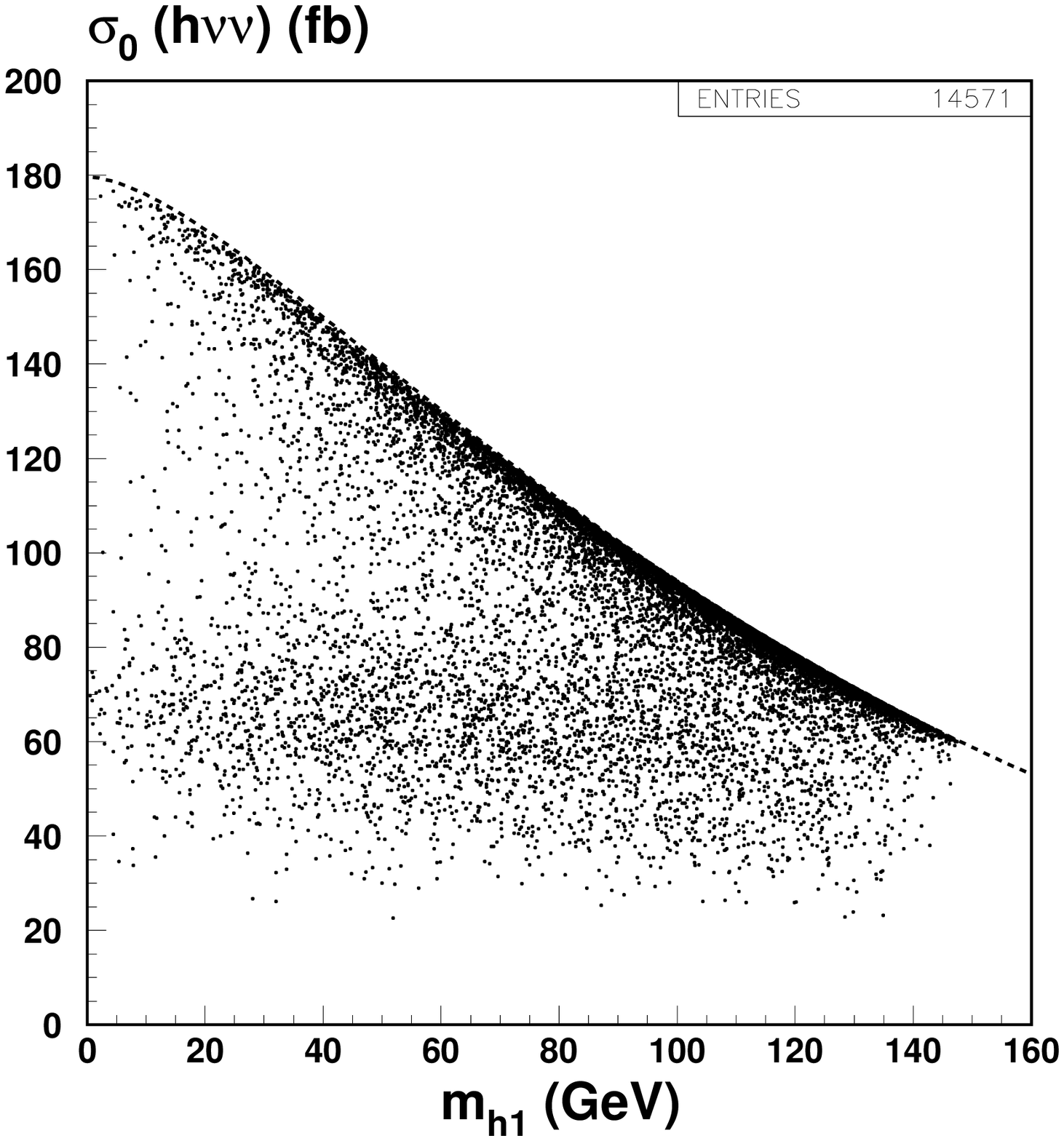}
\caption[plot]{The plot of $\sigma_0 (h \nu {\bar \nu})$ against $m_{h_1}$ in $e^+e^-$ collisions with $\sqrt{s} =500$ GeV, for the same parameter ranges as Fig. 2a.
The dashed curve represents the production cross section of the SM Higgs boson via the $WW$ fusion process.
The minimum value of $\sigma_0 (h \nu {\bar \nu})$ is about 15 fb, represented by a point in the lower right corner at about $m_{h_1} \sim 127$ GeV.}
\end{center}
\end{figure}

\renewcommand\thefigure{2c}
\begin{figure}[t]
\begin{center}
\includegraphics[scale=0.6]{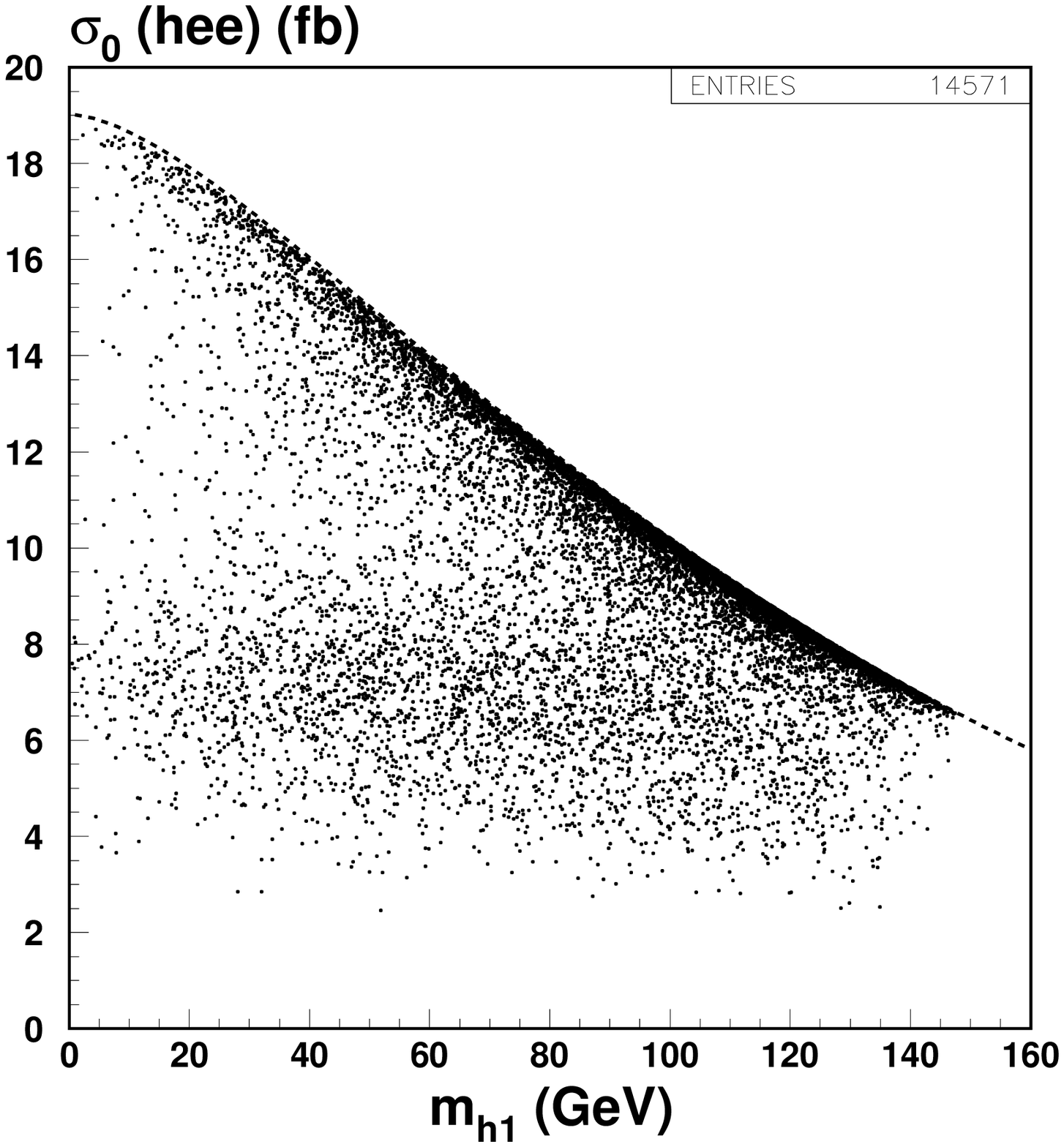}
\caption[plot]{The plot of $\sigma_0 (h e^+e^-)$ against $m_{h_1}$ in $e^+e^-$ collisions with $\sqrt{s} =500$ GeV, for the same parameter ranges as Fig. 2a.
The dashed curve represents the production cross section of the SM Higgs boson via the $ZZ$ fusion process.
The minimum value of $\sigma_0 (h e^+e^-)$ is about 1.5 fb, represented by a point in the lower right corner at about $m_{h_1} \sim 127$ GeV.}
\end{center}
\end{figure}

\renewcommand\thefigure{3a}
\begin{figure}[t]
\begin{center}
\includegraphics[scale=0.6]{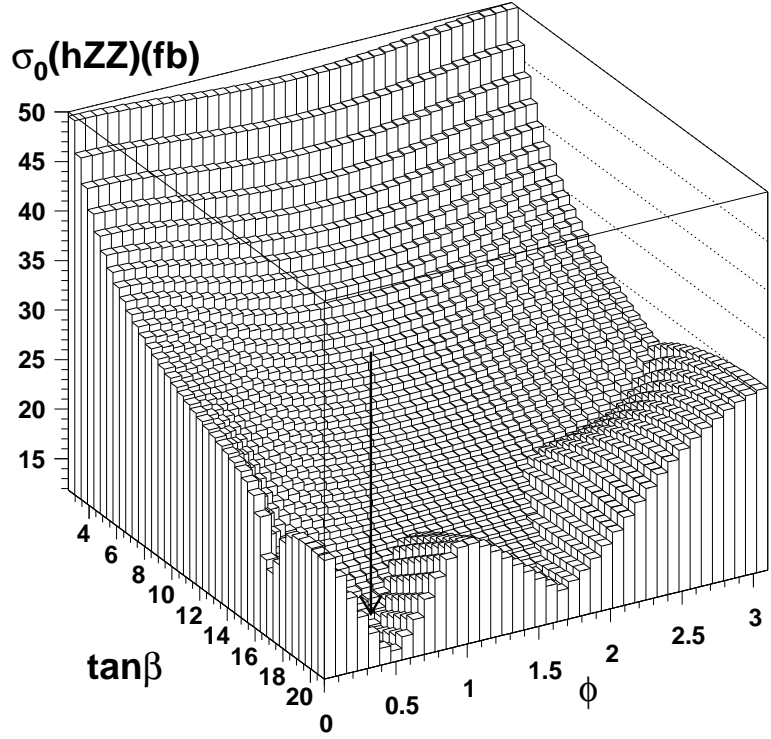}
\caption[plot]{The lego plot of $\sigma_0 (hZZ)$ in the ($\tan \beta, \phi$)-plane
for the Higgs production via the Higgs-strahlung process in $e^+e^-$ collisions
with $\sqrt{s} =500$ GeV, for $\phi_t =\phi$, $m_Q = 766$ GeV, $\lambda = 0.088$,
$k = 0.12$, $A_k = 86$ GeV, $x = 699$ GeV, $A_t = 1350$ GeV, and $m_A = 139$ GeV.
The minimum value of $\sigma_0 (hZZ)$ is about 12 fb, which occurs at a dip
in the forefront of the ($\tan \beta, \phi$)-plane at ($\tan \beta, \phi$) = ($20, 3/\pi$),
indicated by an arrow.}
\end{center}
\end{figure}

\renewcommand\thefigure{3b}
\begin{figure}[t]
\begin{center}
\includegraphics[scale=0.6]{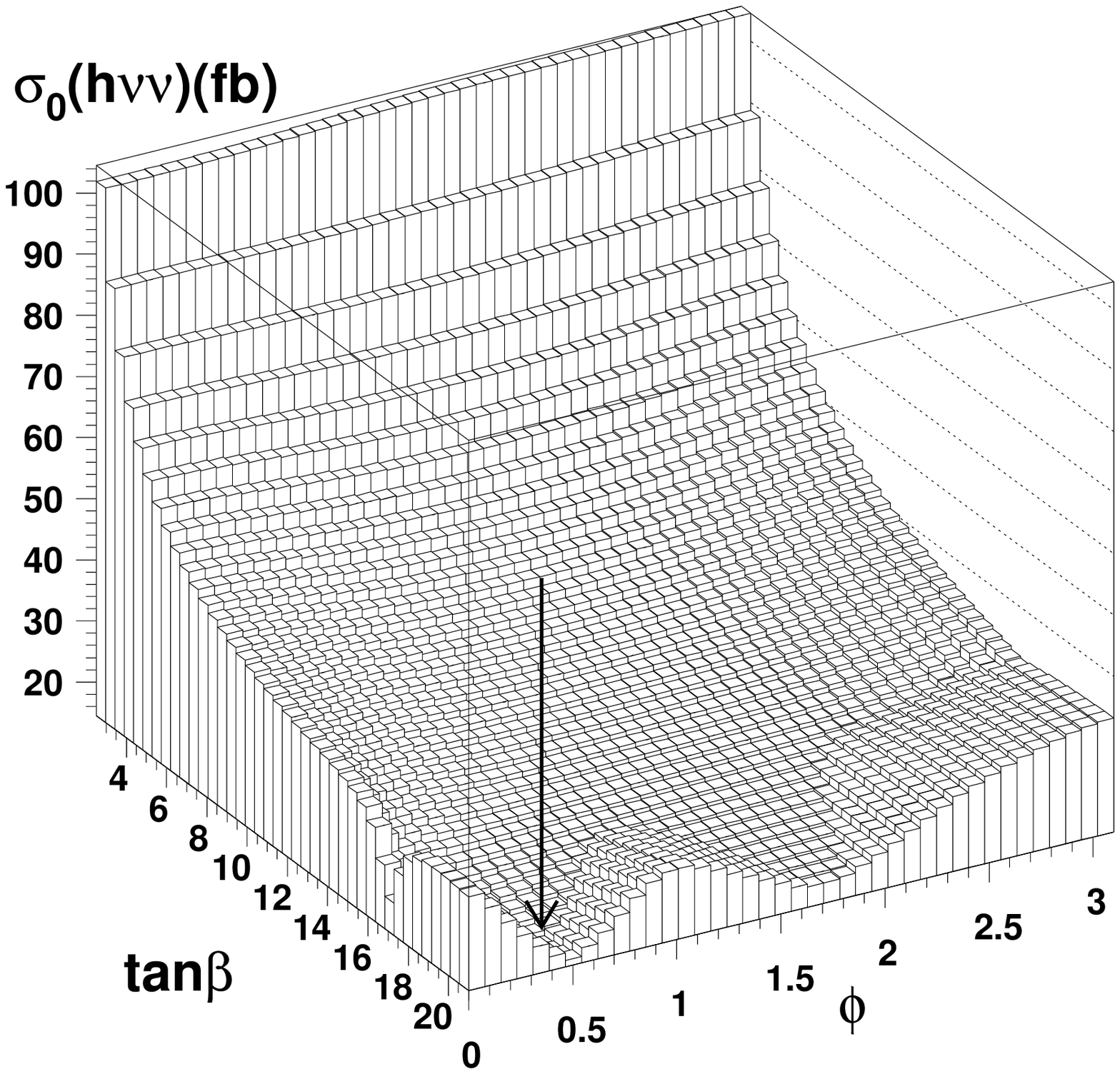}
\caption[plot]{The lego plot of $\sigma_0 (h\nu{\bar \nu})$ in the ($\tan \beta, \phi$)-plane for the Higgs production via the $WW$ fusion process in $e^+e^-$ collisions with $\sqrt{s} =500$ GeV, for the same parameters as Fig. 3a.
The minimum value of $\sigma_0 (h\nu{\bar \nu})$ is about 15 fb, which occurs at the same point of the ($\tan \beta, \phi$)-plane as Fig. 3a.}
\end{center}
\end{figure}

\renewcommand\thefigure{3c}
\begin{figure}[t]
\begin{center}
\includegraphics[scale=0.6]{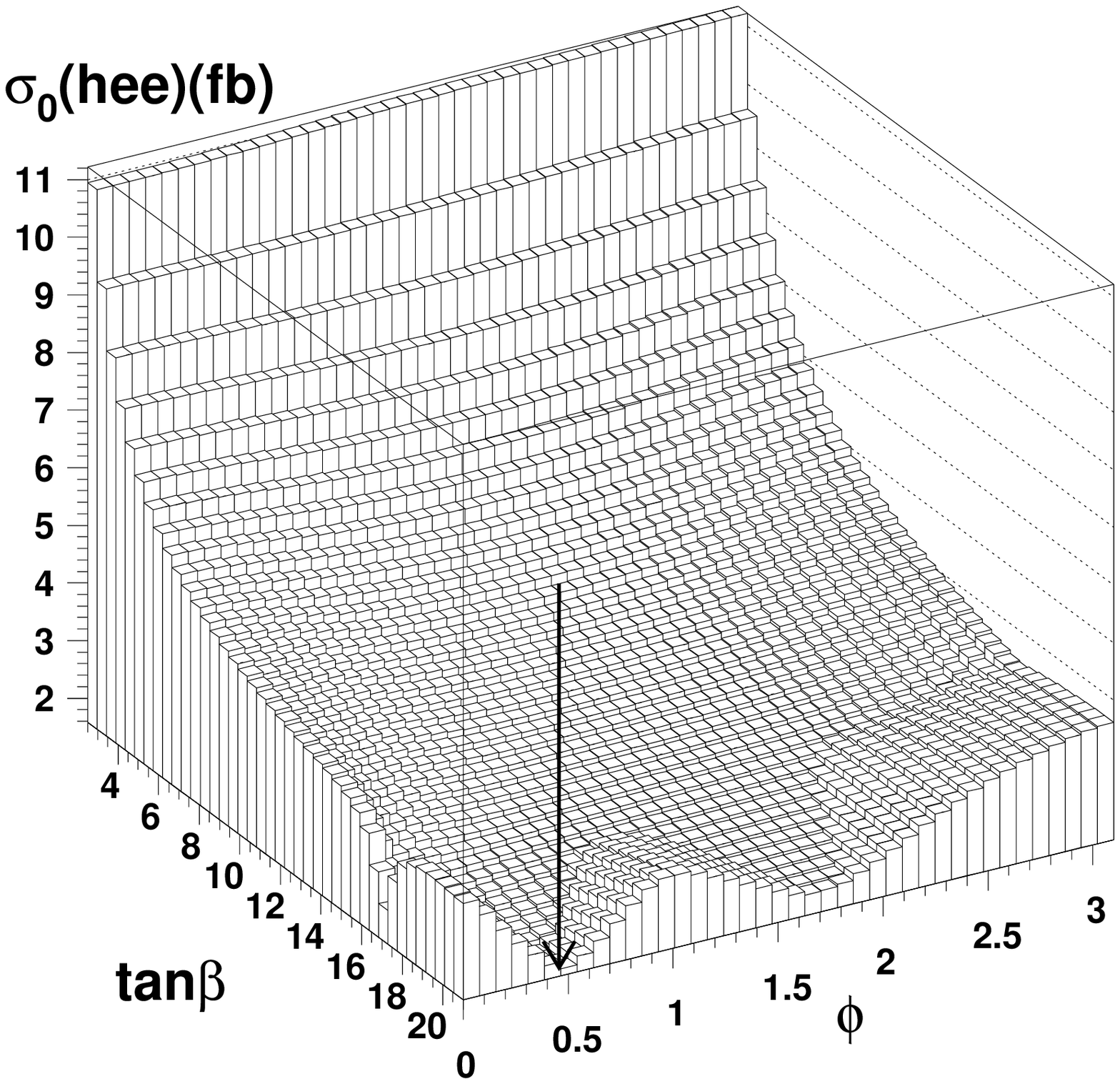}
\caption[plot]{The lego plot of $\sigma_0 (h e^+e^-)$ in the ($\tan \beta, \phi$)-plane for the Higgs production via the $ZZ$ fusion process in $e^+e^-$ collisions with $\sqrt{s} =500$ GeV, for the same parameters as Fig. 3a.
The minimum value of $\sigma_0 (h e^+e^-)$ is about 1.5 fb, which occurs at the same point of the ($\tan \beta, \phi$)-plane as Fig. 3a.}
\end{center}
\end{figure}

\renewcommand\thefigure{4a}
\begin{figure}[t]
\begin{center}
\includegraphics[scale=0.6]{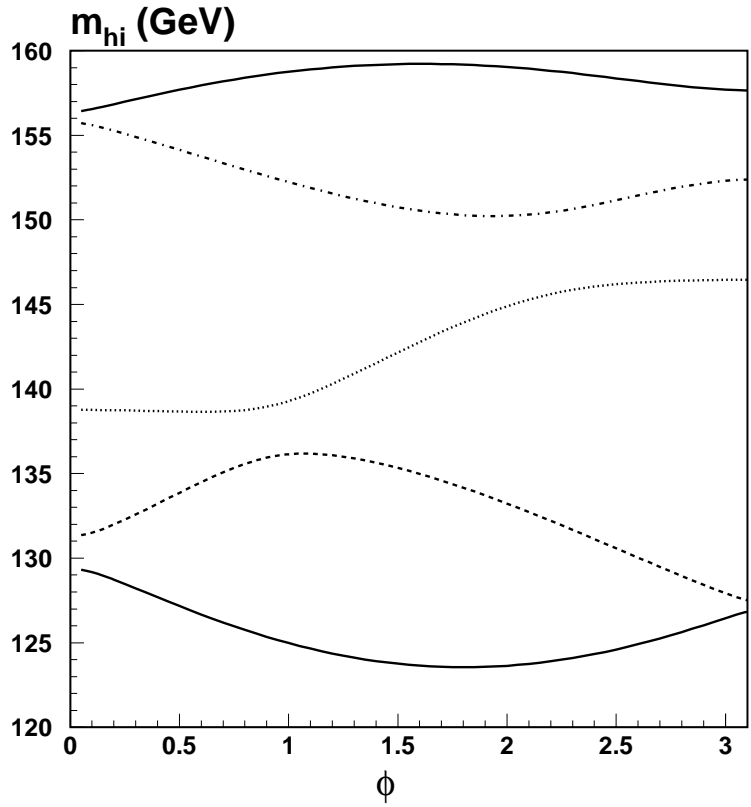}
\caption[plot]{The masses of the five neutral Higgs bosons are plotted as functions of $\phi$, where the parameters are set as $\phi_t =\phi$, $\tan \beta =20$, $m_Q = 766$ GeV, $\lambda = 0.088$, $k = 0.12$, $A_k = 86$ GeV, $x = 699$ GeV, $A_t = 1350$ GeV, and $m_A = 139$ GeV.}
\end{center}
\end{figure}

\renewcommand\thefigure{4b}
\begin{figure}[t]
\begin{center}
\includegraphics[scale=0.6]{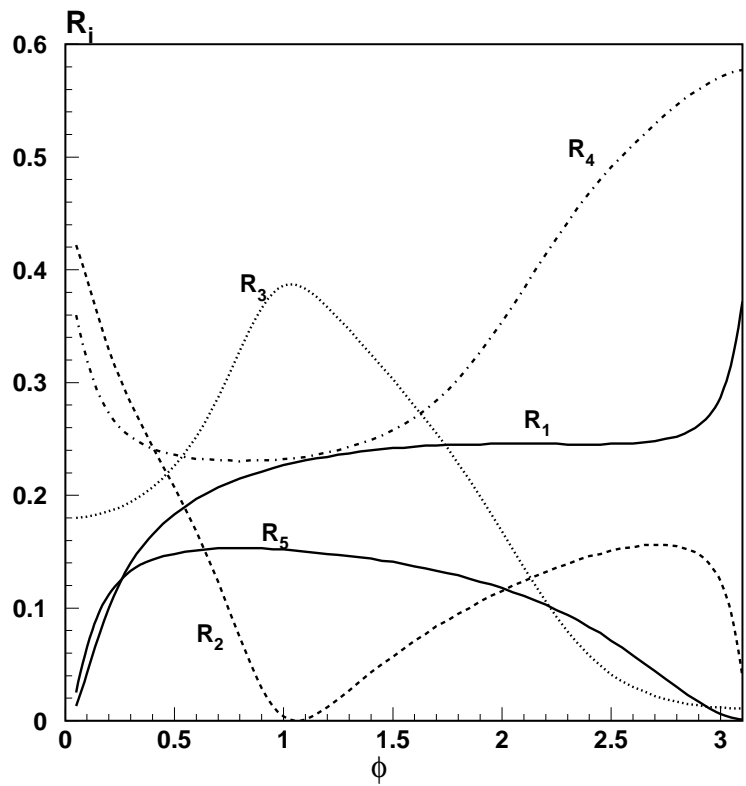}
\caption[plot]{$R_i$ ($i$ = 1-5) as functions of $\phi$, for the same parameters as Fig. 4a.}
\end{center}
\end{figure}

\renewcommand\thefigure{4c}
\begin{figure}[t]
\begin{center}
\includegraphics[scale=0.6]{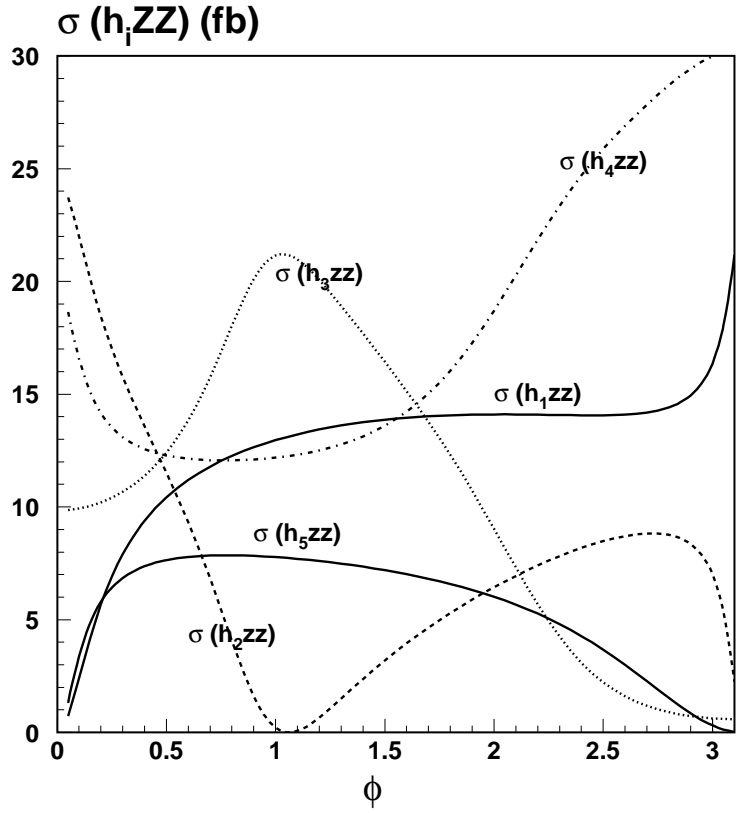}
\caption[plot]{The production cross sections of $h_i$ ($i$ = 1-5) via the Higgs-strahlung process in $e^+e^-$ collisions with $\sqrt{s} =500$ GeV, as functions of $\phi$, for the same parameters as Fig. 4a.}
\end{center}
\end{figure}

\renewcommand\thefigure{4d}
\begin{figure}[t]
\begin{center}
\includegraphics[scale=0.6]{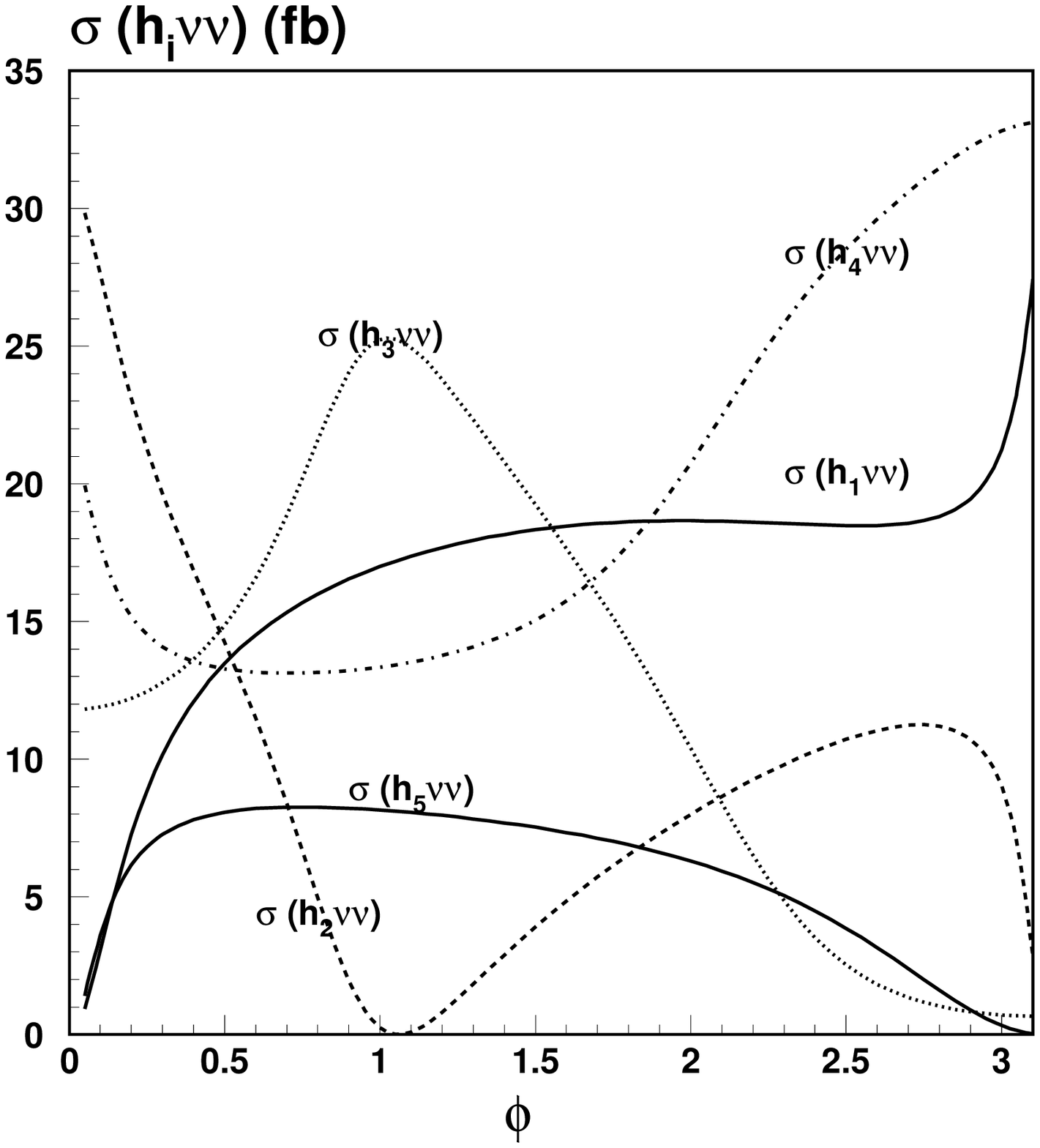}
\caption[plot] {The production cross sections of $h_i$ ($i$ = 1-5) via the $WW$ fusion process in $e^+e^-$ collisions with $\sqrt{s} =500$ GeV, as functions of $\phi$, for the same parameters as Fig. 4a.}
\end{center}
\end{figure}

\renewcommand\thefigure{4e}
\begin{figure}[t]
\begin{center}
\includegraphics[scale=0.6]{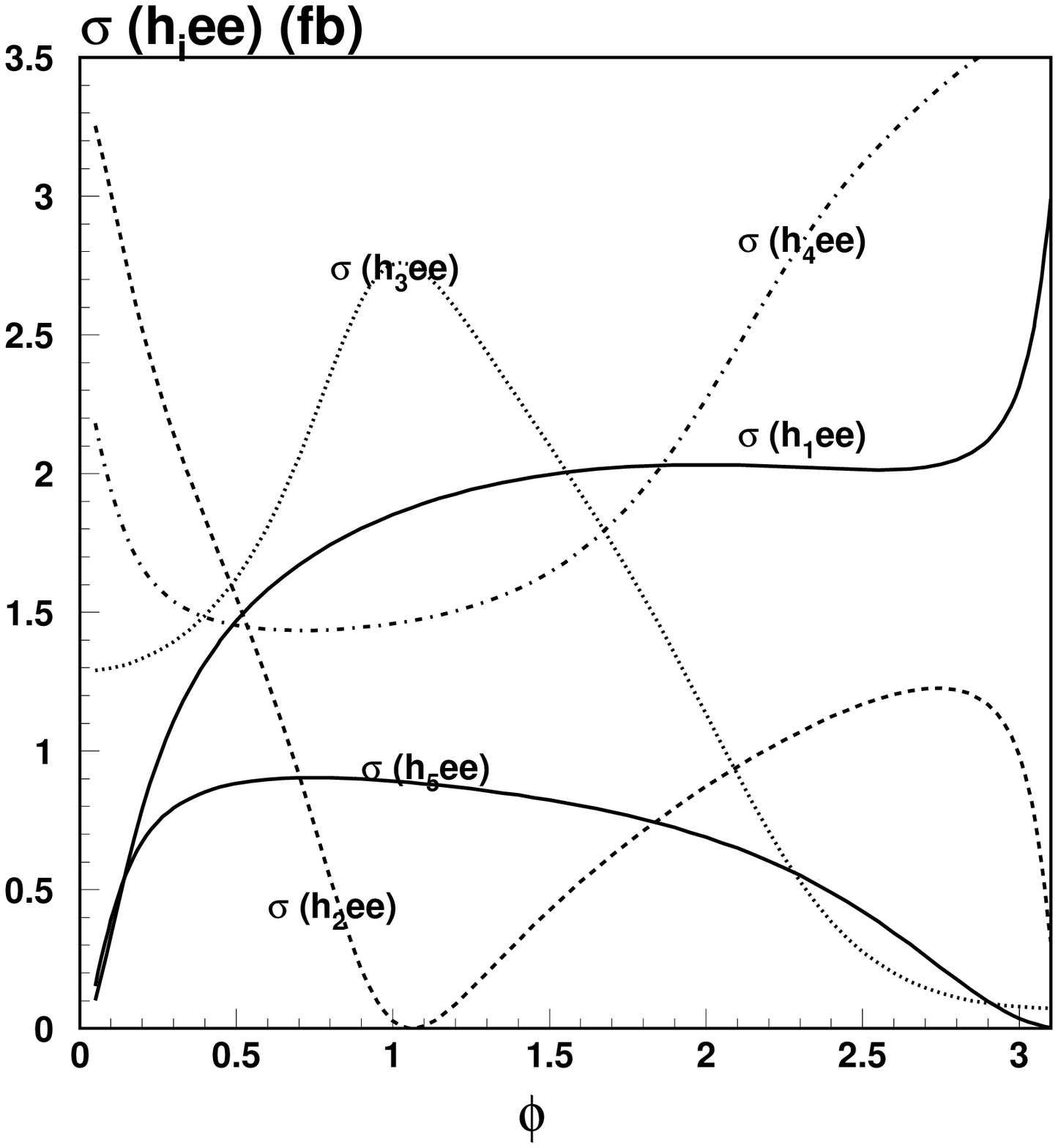}
\caption[plot]{The production cross sections of $h_i$ ($i$ = 1-5) via the $ZZ$ fusion process in $e^+e^-$ collisions with $\sqrt{s} =500$ GeV, as functions of $\phi$, for the same parameters as Fig. 4a.}
\end{center}
\end{figure}

\renewcommand\thefigure{5a}
\begin{figure}[t]
\begin{center}
\includegraphics[scale=0.6]{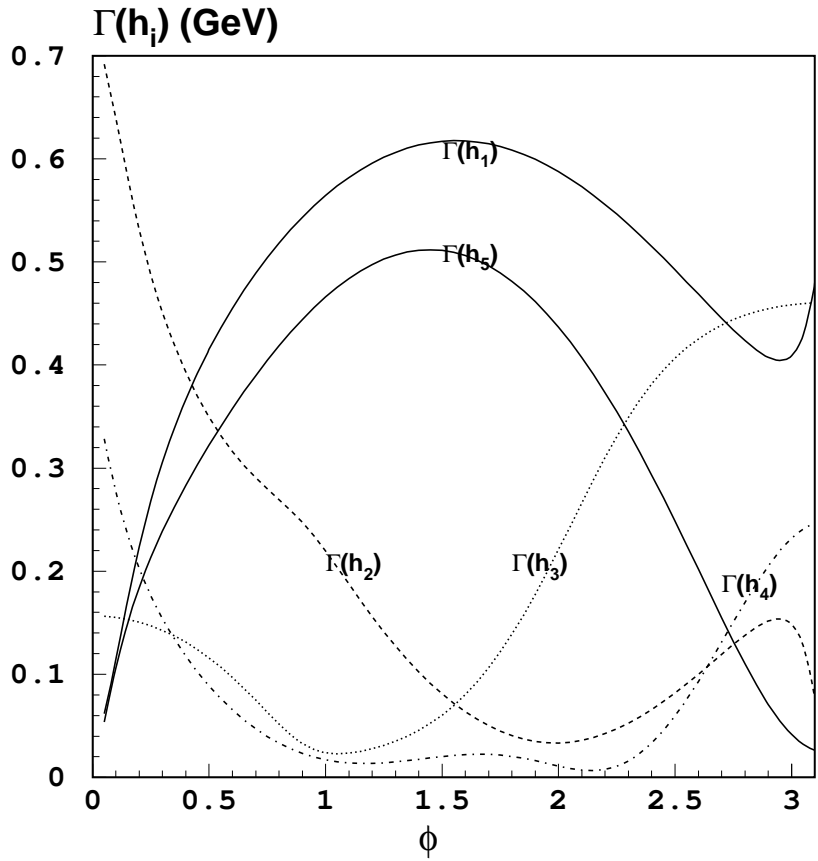}
\caption[plot]{The total decay widths of the five neutral Higgs bosons as functions of $\phi$, for the same parameters as Fig. 4a.}
\end{center}
\end{figure}

\renewcommand\thefigure{5b}
\begin{figure}[t]
\begin{center}
\includegraphics[scale=0.6]{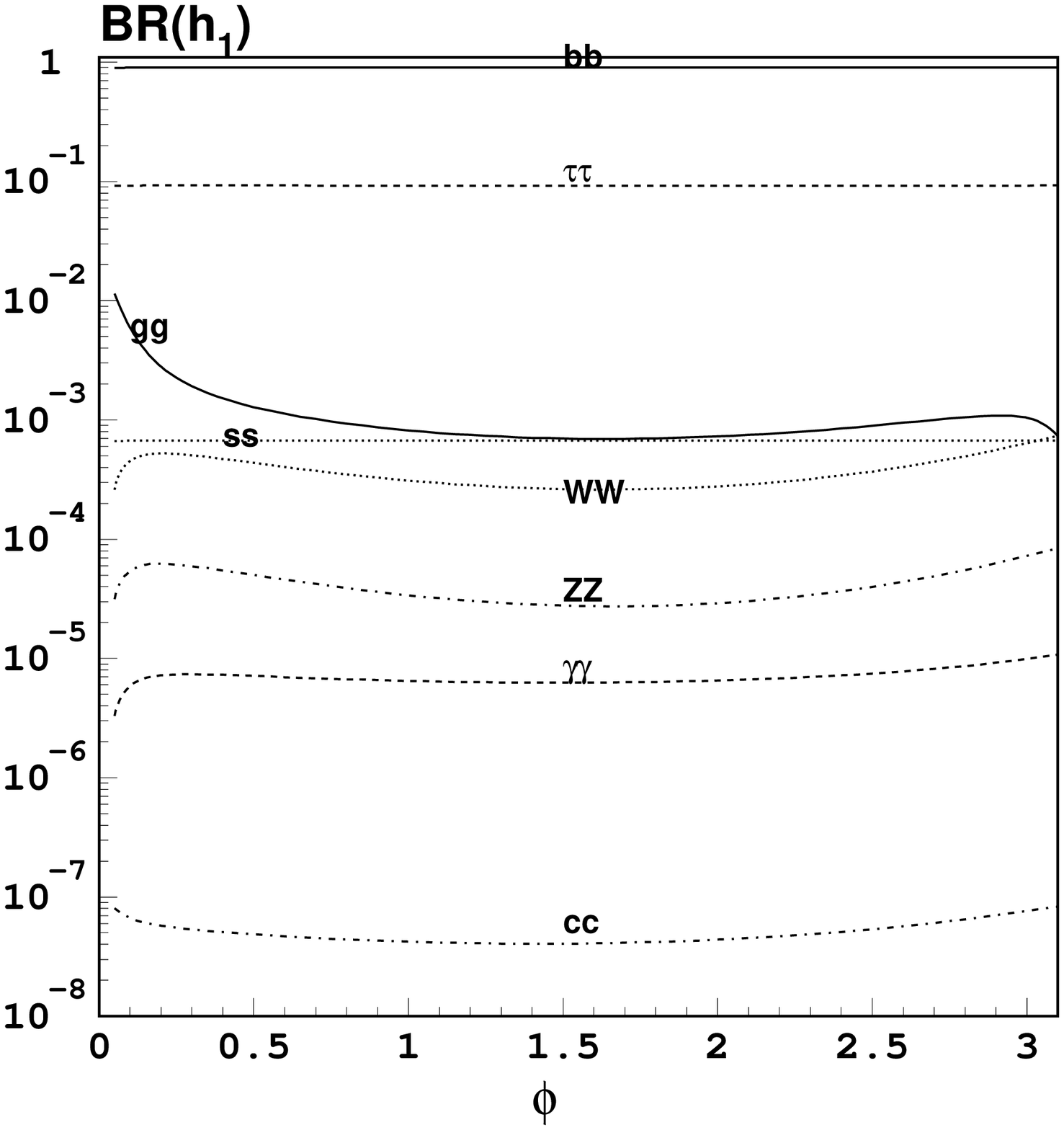}
\caption[plot]{The branching ratios of $h_1$ as functions of $\phi$, for the same parameters as Fig. 4a.}
\end{center}
\end{figure}

\renewcommand\thefigure{5c}
\begin{figure}[t]
\begin{center}
\includegraphics[scale=0.6]{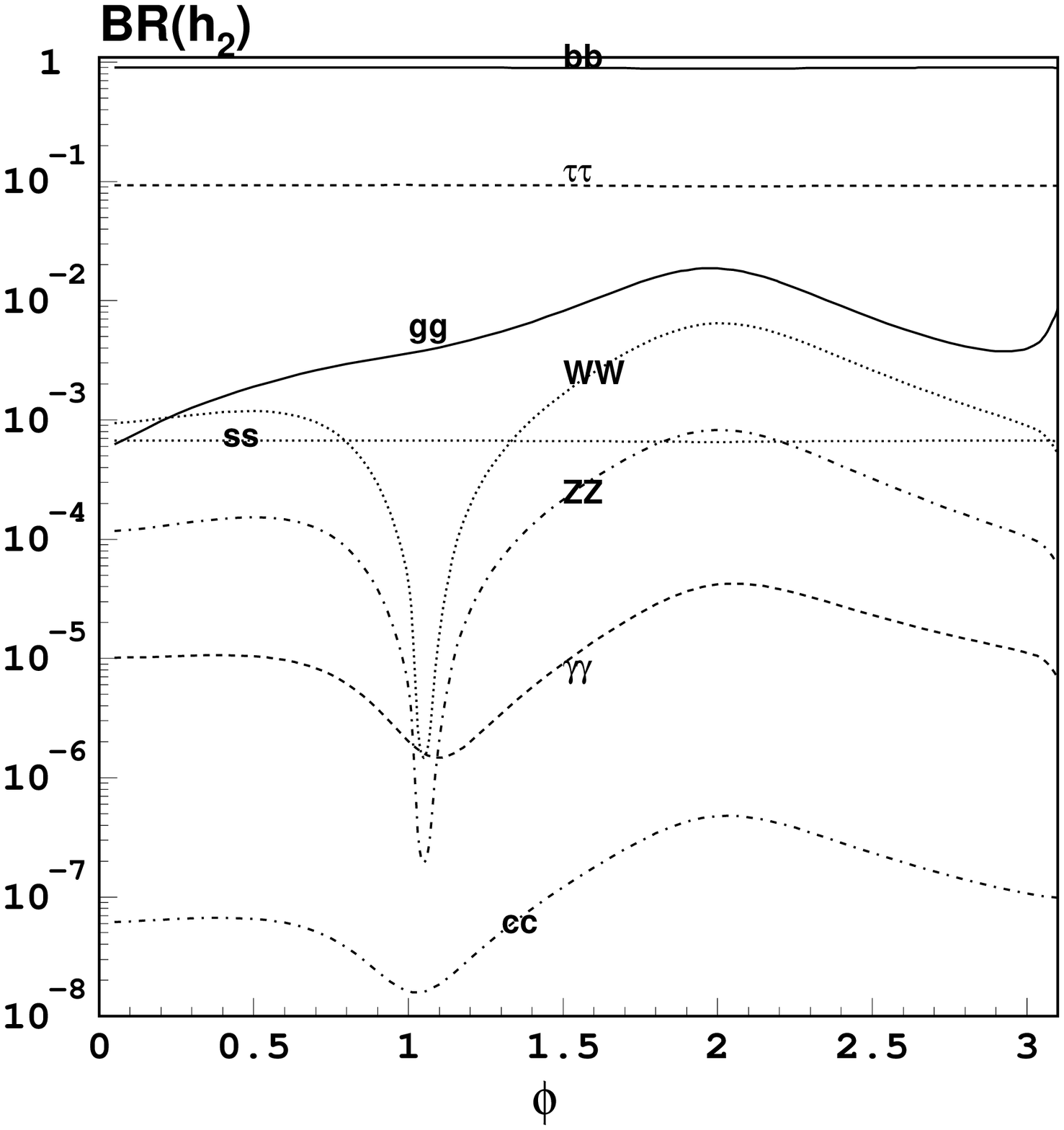}
\caption[plot]{The branching ratios of $h_2$ as functions of $\phi$, for the same parameters as Fig. 4a.}
\end{center}
\end{figure}

\renewcommand\thefigure{5d}
\begin{figure}[t]
\begin{center}
\includegraphics[scale=0.6]{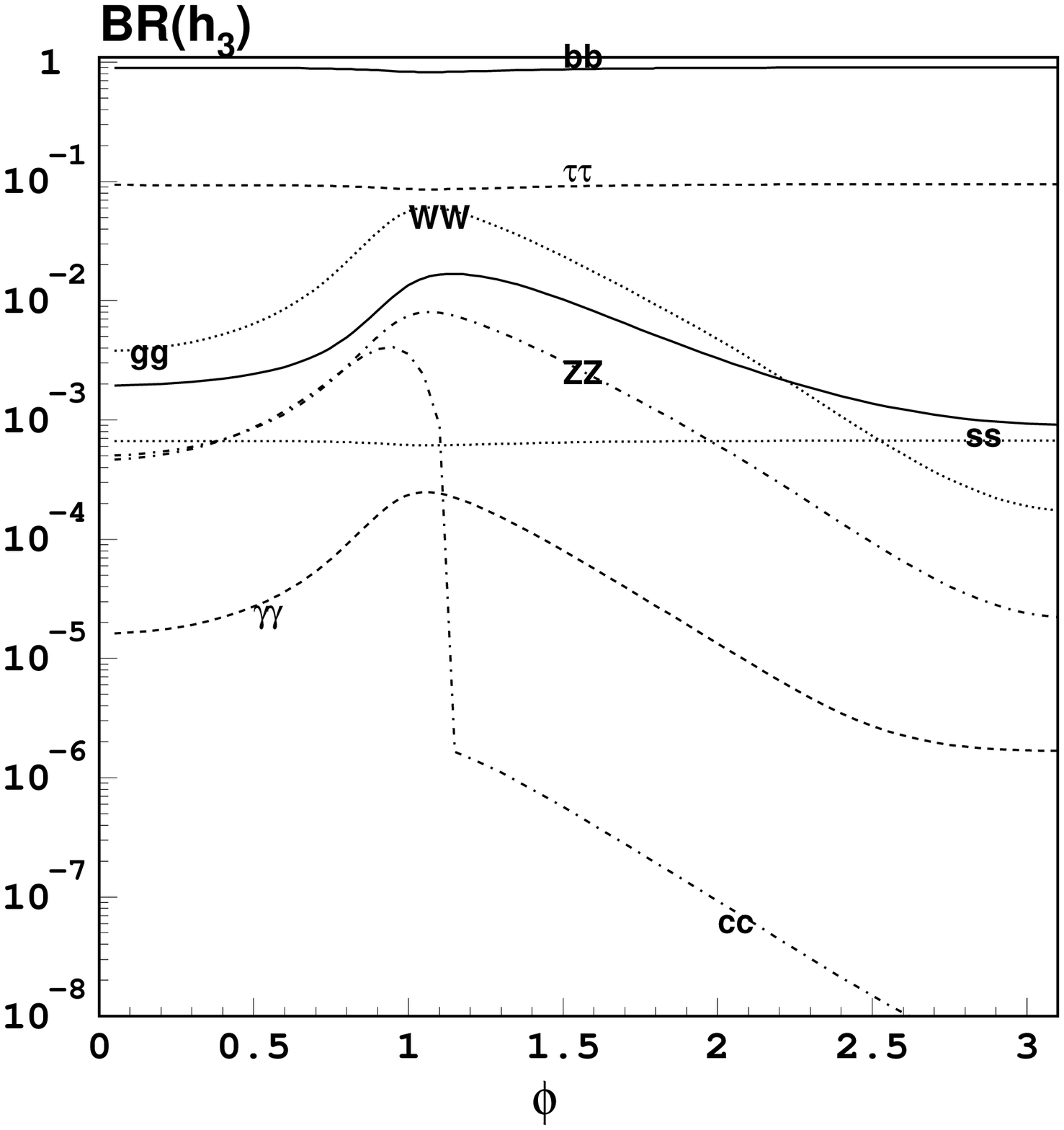}
\caption[plot]{The branching ratios of $h_3$ as functions of $\phi$, for the same parameters as Fig. 4a.}
\end{center}
\end{figure}

\renewcommand\thefigure{5e}
\begin{figure}[t]
\begin{center}
\includegraphics[scale=0.6]{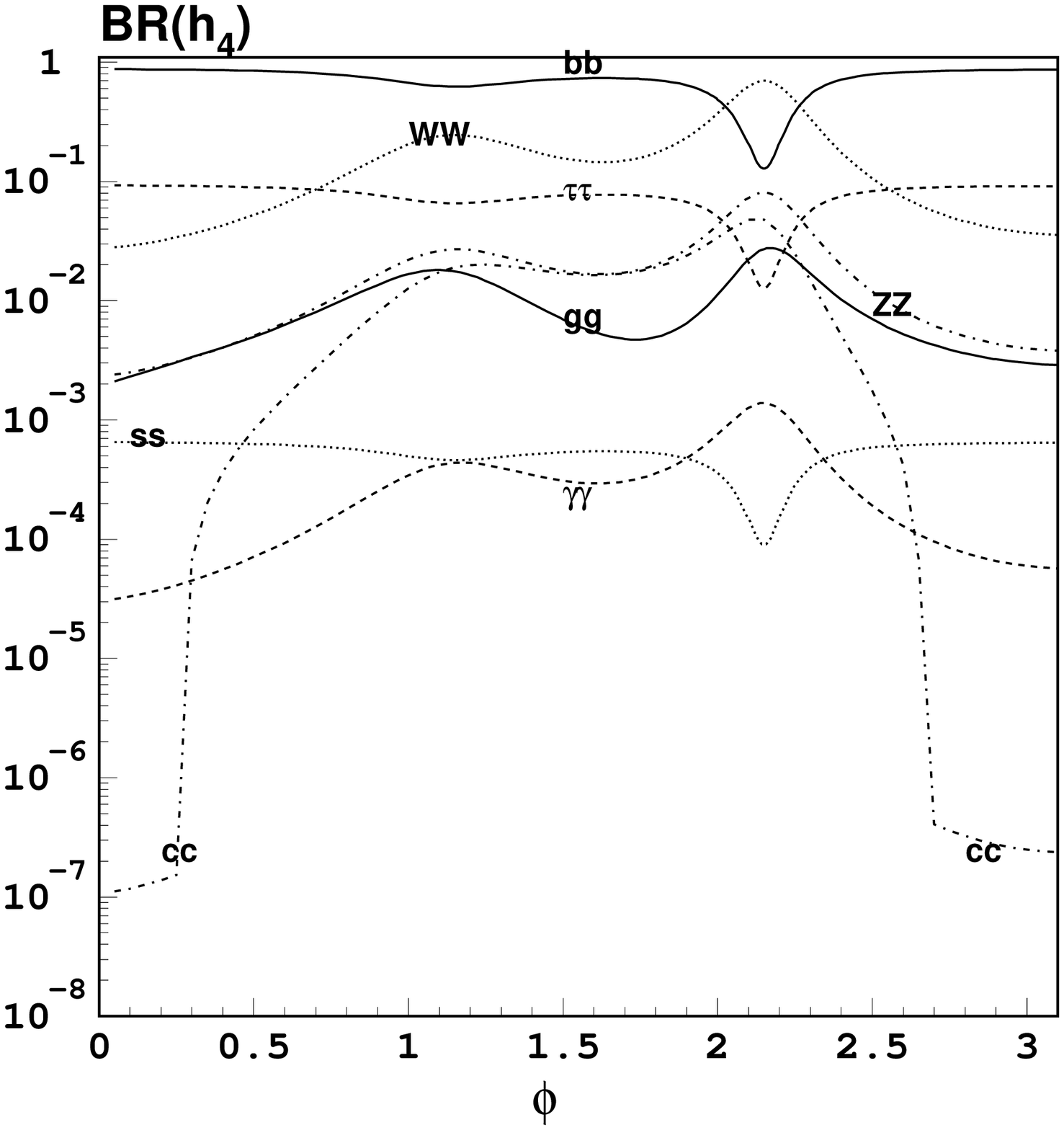}
\caption[plot]{The branching ratios of $h_4$ as functions of $\phi$, for the same parameters as Fig. 4a.}

\end{center}
\end{figure}

\renewcommand\thefigure{5f}
\begin{figure}[t]
\begin{center}
\includegraphics[scale=0.6]{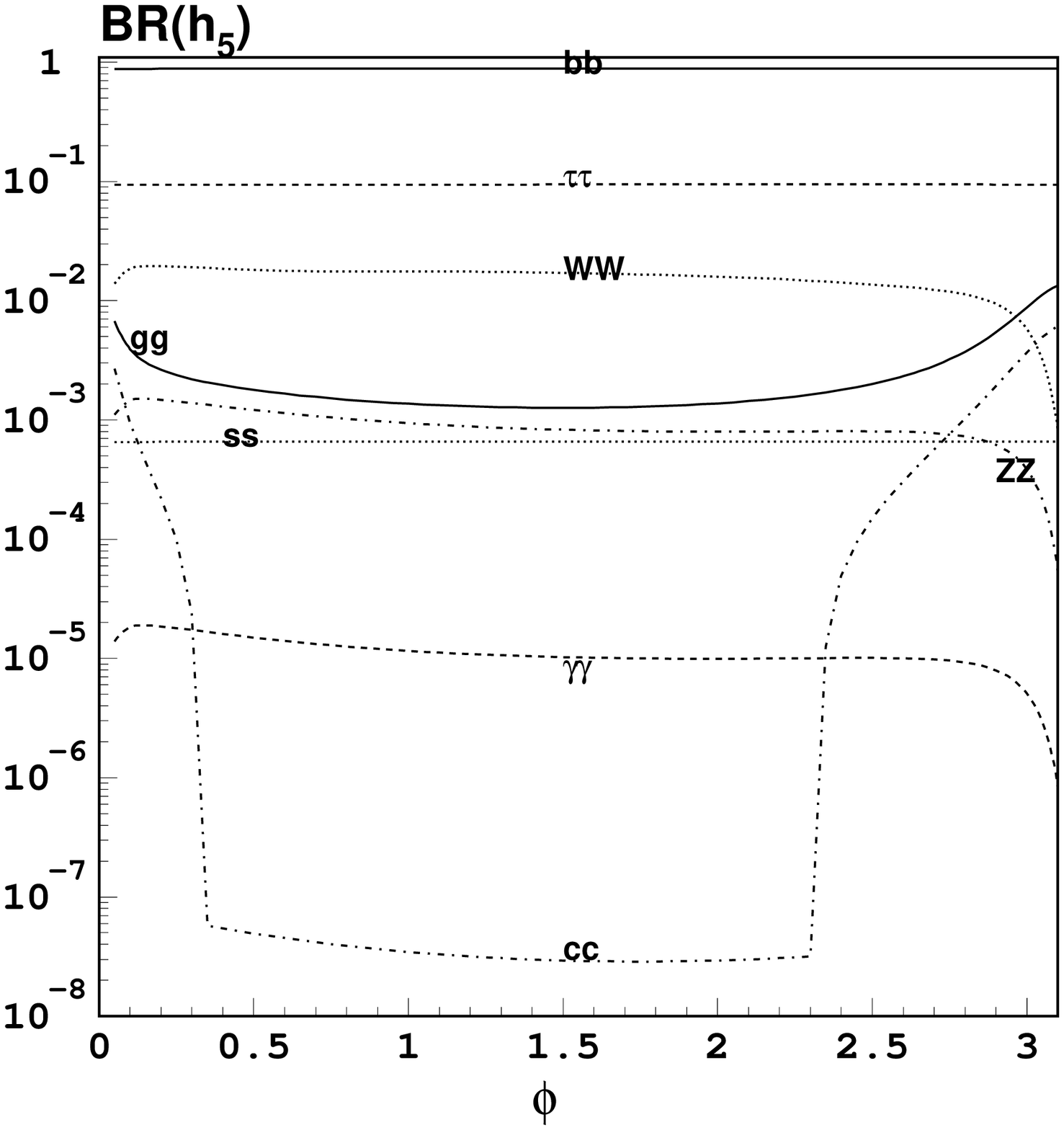}
\caption[plot]{The branching ratios of $h_5$ as functions of $\phi$, for the same parameters as Fig. 4a.}
\end{center}
\end{figure}

\end{document}